\documentclass[11pt]{article}
\usepackage{geometry} 
\usepackage[english]{babel}
\usepackage[english]{layout}
\usepackage{cite}

\usepackage{color}
\usepackage{mathpazo} 
\usepackage[scaled]{helvet} 
\usepackage{courier} 
\normalfont
\usepackage[T1]{fontenc}

\usepackage{amsfonts,latexsym,amsthm,amssymb}
\usepackage{amsmath,bbm}  

\def\fr{{\mathfrak r}}
\def\fs{{\mathfrak s}}

\newcommand{\beq}{\begin{equation}}  
\newcommand{\eeq}{\end{equation}} 
\newcommand{\bea}{\begin{eqnarray}}  
\newcommand{\eea}{\end{eqnarray}} 
\newcommand{\beano}{\begin{eqnarray*}}  
\newcommand{\eeano}{\end{eqnarray*}} 
  
\newcommand{\mb}[1]{\quad\mbox{#1}\quad}

\usepackage{memhfixc}  
\usepackage{fancyhdr} 
\usepackage{sectsty}
\allsectionsfont{\sffamily\mdseries\upshape} 

\numberwithin{equation}{section}

\begin{document}

\title{\LARGE \textbf{Algebraic Bethe ansatz for the $s\ell(2)$ Gaudin model with boundary}}

\author{\textsf{N. ~Cirilo ~Ant\'onio,}
\thanks{E-mail address: nantonio@math.ist.utl.pt}
\textsf{ ~ N. ~Manojlovi\'c,}
\thanks{E-mail address: nmanoj@ualg.pt}
\textsf{ ~ E. ~Ragoucy}
\thanks{E-mail address: eric.ragoucy@lapth.cnrs.fr}
\textsf{~ and I. ~Salom}
\thanks{E-mail address: isalom@ipb.ac.rs} \\
\\
\textit{$^{\ast}$Centro de An\'alise Funcional e Aplica\c{c}\~oes}\\
\textit{Instituto Superior T\'ecnico, Universidade de Lisboa} \\
\textit{Av. Rovisco Pais, 1049-001 Lisboa, Portugal} \\
\\
\textit{$^{\dag}$Grupo de F\'{\i}sica Matem\'atica da Universidade de Lisboa} \\
\textit{Av. Prof. Gama Pinto 2, PT-1649-003 Lisboa, Portugal} \\
\\
\textit{$^{\dag}$Departamento de Matem\'atica, F. C. T.,
Universidade do Algarve} \\
\textit{Campus de Gambelas, PT-8005-139 Faro, Portugal}\\
\\
\textit{$^{\ddag}$Laboratoire d'Annecy-le-Vieux de Physique Th\'eorique LAPTh}\\
\textit{CNRS et Univerit\'e de Savoie, UMR 5108, B.P. 110}\\ 
\textit{74941 Annecy-le-Vieux Cedex, France}\\
\\
\textit{$^{\S}$Institute of Physics, University of Belgrade}\\
\textit{P.O. Box 57, 11080 Belgrade, Serbia}\\
\\
}

\date{}

\maketitle
\thispagestyle{empty}

\begin{abstract}
Following Sklyanin's proposal in the periodic case, we derive the generating function of the Gaudin Hamiltonians with boundary terms. Our derivation is based on the quasi-classical expansion of the linear combination of the transfer matrix of the XXX Heisenberg spin chain and the central element, the so-called Sklyanin determinant. The corresponding Gaudin Hamiltonians with boundary terms are obtained as the residues of the generating function. By defining the appropriate Bethe vectors which yield strikingly simple off shell action of the generating function, we fully implement the algebraic Bethe ansatz, obtaining the spectrum of the generating function and the corresponding Bethe equations.
\end{abstract}

\clearpage
\newpage

\section{Introduction}
A model of interacting spins in a chain was first considered by Gaudin \cite{Gaudin76,Gaudin83}. Gaudin derived these models as a quasi-classical limit of the quantum chains. 
Sklyanin studied the rational $s\ell(2)$ model in the framework of the quantum inverse scattering method using the $s\ell(2)$ invariant classical r-matrix \cite{Sklyanin89}. A generalization of these results to all cases when skew-symmetric r-matrix satisfies the classical Yang-Baxter equation \cite{BelavinDrinfeld} was relatively straightforward \cite{SklyaninTakebe,Semenov97}. Therefore, considerable attention has been devoted to Gaudin models corresponding to the the classical r-matrices of simple Lie algebras \cite{Jurco89, Jurco90, BabujianFlume, FeiginFrenkelReshetikhin, ReshetikhinVarchenko, WagnerMacfarlane00} and Lie superalgebras \cite{BrzezinskiMacfarlane94, KulishManojlovic01, KulishManojlovic03, LimaUtiel01, KurakLima04}. 

Hikami, Kulish and Wadati showed that the quasi-classical expansion of the  transfer matrix of the periodic chain, calculated at the special values of the spectral parameter, yields the Gaudin Hamiltonians \cite{HikamiKulishWadati92,HikamiKulishWadati92a}. Hikami showed how the quasi-classical expansion of the transfer matrix, calculated at the special values of the spectral parameter, yields the Gaudin Hamiltonians in the case of non-periodic boundary conditions \cite{Hikami95}. Then the ABA was applied to open Gaudin model in the context of the the Vertex-IRF correspondence \cite{YangZhangSasakic04, YangZhangSasakic05, YangZhang12}. Also,  results were obtained for the open Gaudin models based on Lie superalgebras \cite{Lima09}. An approach to study the open Gaudin models based on the classical reflection equation \cite{Sklyanin86,Sklyanin87,Sklyanin88}
and the non-unitary r-matrices was developed recently, see \cite{Skrypnyk09, Skrypnyk13} and the references therein. For a review of the open Gaudin model see \cite{CAMN}. Progress in applying Bethe ansatz to the Heisenberg spin chain with non-periodic boundary conditions compatible with the integrability of the quantum systems \cite{China03, Nepomechie04, French06, Martins05, YangWang13, YangWang13a, Eric2013RMP, Eric13, Belliard13, Lima13,CAMS} had recent impact on the study of the corresponding Gaudin model \cite{CAMS,HaoCaoYang14}. The so-called $T-Q$ approach to implementation of  Bethe ansatz \cite{YangWang13, YangWang13a} was used to obtain the eigenvalues of the associated Gaudin Hamiltonians and the corresponding Bethe ansatz equations \cite{HaoCaoYang14}. In \cite{CAMS} the off shell action of the generating function of the Gaudin Hamiltonians on the Bethe vectors was obtained through the so-called quasi-classical limit.

Here we derive the generating function of the Gaudin Hamiltonians with boundary terms following Sklyanin's approach in the periodic case  \cite{Sklyanin89}. Our derivation is based on the quasi-classical expansion of the linear combination of the transfer matrix of the inhomogeneous XXX Heisenberg spin chain and the central element, the so-called Sklyanin determinant. The essential step in this derivation is the expansion of the monodromy matrix in powers of the quasi-classical parameter. Moreover, we show how the representation of the relevant Lax matrix in terms of local spin operators yields the partial fraction decomposition of the generating function. Consequently, the Gaudin Hamiltonians with the boundary terms are obtained from the residues of the generating function at poles. We derive the relevant linear bracket  for the Gaudin Lax operator and certain classical r-matrix, obtained form the $s\ell(2)$ invariant classical r-matrix and the corresponding K-matrix. The local realisation of the Lax matrix together with the linear bracket provide the necessary structure for the implementation of the algebraic Bethe ansatz. In this framework, the Bethe vectors, defined as the symmetric functions of its arguments, have a remarkable property that the off shell action of the generating function on them is strikingly simple. Actually, it is as simple as it can be since it  practically coincide with the corresponding formula in the case when the boundary matrix is diagonal \cite{Hikami95}. The off shell action of the generating function of the Gaudin Hamiltonians with the boundary terms yields the spectrum of the system and the corresponding Bethe equations. As usual, when the Bethe equations are imposed on the parameters of the Bethe vectors, the unwanted terms in the action of the generating function are annihilated. 

However, more compact form of the Bethe vector $\varphi _M (\mu _1, \mu _2, \dots , \mu _M)$, for an arbitrary positive integer $M$, requires further studies. As it is evident form the formulas for the Bethe vector $\varphi _4 (\mu _1, \mu _2, \mu _3, \mu _4)$ given in the Appendix B, the problem lies in the definition the scalar coefficients $c_M^{(m)}(\mu _1 , \dots \mu _m ; \mu _{m+1}, \dots, \mu _M)$, with $m = 1, 2, \dots , M$. Some of them are straightforward to obtain but, in particular, the coefficient   $c_M^{(M)}(\mu _1 , \mu _2 \dots , \mu _M)$ still represents a challenge, at least in the present form.

This paper is organized as follows. In Section 2 we review the $SL(2)$-invariant Yang R-matrix and provide fundamental tools for the study of the inhomogeneous XXX Heisenberg spin chain and the corresponding Gaudin model. Moreover, we outline Sklyanin's derivation of the rational $s\ell(2)$ Gaudin model. The general solutions of the reflection equation and the dual reflection equation are given in Section 3. As one of the main results of the paper,  the generating function of the Gaudin Hamiltonians with boundary terms is derived in Section 4, using the quasi-classical expansion of the linear combination of the transfer matrix of the inhomogeneous XXX spin chain and the so-called Sklyanin determinant. The relevant algebraic structure, including the classical reflection equation, is given in Section 5. The implementation of the algebraic Bethe ansatz is presented in Section 6, including the definition of the Bethe vectors and the formulae of the off shell action of the generating function of the Gaudin Hamiltonians. Our conclusions are presented in the Section 7. Finally, in Appendix A are given some basic definitions for the convenience of the reader.

\section{$s\ell(2)$ Gaudin model}
The XXX Heisenberg spin chain is related to the $SL(2)$-invariant Yang R-matrix \cite{Yang67}
\begin{equation}
\label{YangR}
R (\lambda) = \lambda \mathbbm{1}  + \eta \mathcal{P } = \left(\begin{array}{cccc}
\lambda + \eta & 0 & 0 & 0 \\
0 & \lambda & \eta & 0 \\
0 & \eta & \lambda  & 0 \\
0 & 0 & 0 & \lambda + \eta \end{array}\right),
\end{equation}
where $\lambda$ is a spectral parameter, $\eta$ is a quasi-classical parameter, $\mathbbm{1}$ is the identity operator and we use $\mathcal{P }$ for the permutation in $\mathbb{C} ^2 \otimes \mathbb{C} ^2$.

The Yang R-matrix satisfies the Yang-Baxter equation  \cite {Yang67, Baxter82, Faddeev98, Kulish09} in the space $\mathbb{C}^2 \otimes \mathbb{C}^2 \otimes \mathbb{C}^2$
\begin{equation}
\label{YBE}
R_{12} ( \lambda - \mu) R_{13} ( \lambda) R_{23} (\mu) = R_{23} (\mu ) R_{13} (\lambda ) R_{12} ( \lambda - \mu),
\end{equation}
we use the standard notation of the quantum inverse scattering method to denote spaces 
on which corresponding $R$-matrices $R_{ij}, ij = 12, 13, 23$ act non-trivially 
and suppress the dependence on the quasi-classical parameter $\eta$ \cite{Faddeev98, Kulish09}.

The Yang R-matrix also satisfies other relevant properties such as

\begin{tabbing}{|l}
xxxxxxxxxxxxxxxxxxxxxxxxxx   \= xxxxxxxxxxxxxxxxxxxxxxxxxx     \kill
unitarity    \>   $R_{12} ( \lambda ) R_{21} ( -\lambda ) = (\eta ^2 - \lambda ^2 )  \mathbbm{1}$;    \\
parity invariance    \> $R_{21} ( \lambda ) = R_{12} ( \lambda ) $;  \\
temporal invariance  \> $R_{12}^t ( \lambda) = R_{12} ( \lambda)$; \\
crossing symmetry  \>  $R ( \lambda) =   \mathcal{J} _1 R ^{t_2}( -\lambda - \eta ) {\mathcal{J}} _1,$
\end{tabbing}
where $t_2$ denotes the transpose in the second space and the entries of the two-by-two matrix $\mathcal{J}$ are $\mathcal{J}_{ab}=  (-1)^{a-1}\delta_{a,3-b}$.

Here we study the inhomogeneous XXX spin chain with $N$ sites, characterised by the local space 
$V_ m = \mathbb{C}^{2s+1}$ and inhomogeneous parameter $\alpha _m$. For simplicity, we start by considering the periodic boundary conditions. The Hilbert space of the system is
\begin{equation}
\label{H-space}
\mathcal{H} = \underset {m=1}{\overset {N}{\otimes}}  V_m = (\mathbb{C}^{2s+1} ) ^{\otimes N}.
\end{equation}
Following \cite{Sklyanin89} we introduce the Lax operator \cite{CAMS}
\begin{equation}
\label{L-XXX}
\mathbb{L}_{0m}(\lambda) 
= \mathbbm{1} + \frac{\eta}{\lambda} \left( \vec{\sigma}_{0} \cdot \vec{S}_{m} \right) 
= \frac{1}{\lambda}  \left(\begin{array}{cc}  
\lambda + \eta S_m^{3} & \eta  S_m^{-} \\ \eta S_m^{+} & \lambda - \eta S_m^{3}
\end{array}\right).
\end{equation}
Notice that $\mathbb{L}(\lambda)$ is a two-by-two matrix in the auxiliary space $V_0 = \mathbb{C}^2$.
It obeys
\begin{equation}
\label{unit:Lax}
\mathbb{L}_{0m}(\lambda) \mathbb{L}_{0m}(\eta-\lambda) 
= \Big(1+\frac{\eta^2\, c_{2,m}}{\lambda(\eta-\lambda)}\Big) \mathbbm{1} _0 \,,
\end{equation}
where $c_{2,m}$ is the value of the Casimir operator on the space $V_m$ \cite{CAMS}.

When the quantum space is also a spin $\frac12$ representation, the Lax operator becomes the $R$-matrix,
$\mathbb{L}_{0m}(\lambda) = \frac{1}{\lambda} R_{0m}\left(\lambda - {\eta}/{2} \right) $.

Due to the commutation relations \eqref{crspin1}, it is straightforward to check that the Lax operator satisfies the RLL-relations
\begin{equation}
\label{RLL}
R_{00'} ( \lambda - \mu) \mathbb{L}_{0m}( \lambda ) \mathbb{L}_{0'm}( \mu ) =  \mathbb{L}_{0'm}( \mu ) \mathbb{L}_{0m}( \lambda )R_{00'} ( \lambda - \mu).
\end{equation}

The so-called monodromy matrix
\begin{equation}
\label{monodromy-T}
T(\lambda ) = \mathbb{L}_{0N} ( \lambda - \alpha _N) \cdots \mathbb{L}_{01} ( \lambda - \alpha _1)
\end{equation}
is used to describe the system. For simplicity we have omitted the dependence on the quasi-classical parameter $\eta$ and the inhomogeneous parameters $\{ \alpha _j , j = 1 , \ldots , N \}$. Notice that $T(\lambda)$ is a two-by-two matrix acting in the auxiliary space $V_0 = \mathbb{C}^2$, whose entries are operators acting in $\mathcal{H}$. From RLL-relations \eqref{RLL} it follows that the monodromy matrix satisfies the RTT-relations
\begin{equation}
\label{RTT}
R_{00'} ( \lambda - \mu) {T}_{0} (\lambda ) {T}_ {0'}(\mu ) =  {T} _ {0'}(\mu ){T}_ {0}(\lambda ) R_{00'} ( \lambda - \mu).
\end{equation}

The periodic boundary conditions and the RTT-relations \eqref{RTT} imply that the transfer matrix
\begin{equation}
\label{periodic-t}
t (\lambda ) = \mathrm{tr}_0 T(\lambda) ,
\end{equation}
commute at different values of the spectral parameter,
\begin{equation}
\label{periodic-tt}
[t (\mu) , t (\nu)] = 0,
\end{equation}
here we have omitted the nonessential arguments.

The RTT-relations admit a central element 
\begin{equation}
\label{DeltaT}
\Delta \left[T(\lambda)\right] = \mathrm{tr}_{00'} P^{-}_{00'} T_{0}\left(\lambda - {\eta}/{2}\right)T_{0'}\left(\lambda + {\eta}/{2}\right),
\end{equation}
where
\begin{equation}
\label{P12min}
P^{-}_{00'} = \frac{\mathbbm{1}  - \mathcal{P}_{00'} }{2} = -  \frac{ 1}{2\eta}\,R_{00'}\!\left(- \eta \right) .
\end{equation}
A straightforward calculation shows that
\begin{equation}
\label{Delta-central}
\Big[ \Delta \big[T(\mu) \big] \, ,\, T(\nu) \Big] = 0.
\end{equation}

As the first step toward the study of the Gaudin model we consider the expansion of the monodromy matrix \eqref{monodromy-T} with respect to the quasi-classical parameter $\eta$
\begin{align}
\label{expan-T}
T(\lambda ) &= \mathbbm{1} + \eta \sum _{m=1}^N \frac{\vec{\sigma}_{0} \cdot \vec{S}_{m}}{\lambda - \alpha _m} + \frac{\eta ^2}{2}
\underset {n \neq m} {\sum _{n,m=1}^N} \frac{\mathbbm{1}_0 \left( \vec{S}_{m} \cdot \vec{S}_{n} \right)}{(\lambda - \alpha _m)(\lambda - \alpha _n)} \notag \\
&+  \frac{\eta ^2}{2} \sum _{m=1}^N \left(
\sum _{n > m}^N  \frac{\imath \vec{\sigma}_{0} \cdot \left( \vec{S}_{n} \times \vec{S}_{m} \right)}{(\lambda - \alpha _m)(\lambda - \alpha _n)} 
+  \sum _ {n<m}^N  \frac{\imath \vec{\sigma}_{0} \cdot \left( \vec{S}_{m} \times \vec{S}_{n} \right)}{(\lambda - \alpha _m)(\lambda - \alpha _n)} \right) 
+ \mathcal{O}(\eta ^3) .
\end{align}
If the Gaudin Lax matrix is defined by \cite{Sklyanin89}
\begin{equation}
\label{L-Gaudin}
L_0(\lambda) = \sum _{m=1}^N \frac{\vec{\sigma}_{0} \cdot \vec{S}_{m}}{\lambda - \alpha _m} 
\end{equation}
and the quasi-classical property of the Yang R-matrix \cite{Sklyanin89}
\begin{equation}
\label{classical-r}
\frac{1}{\lambda} R (\lambda) = \mathbbm{1} - \eta r(\lambda), \qquad \text{where} \qquad r(\lambda) = - \frac{ \mathcal{P}}{\lambda}
\end{equation}
is taken into account, then substitution of the expansion \eqref{expan-T} into the RTT-relations \eqref{RTT} yields the so-called Sklyanin linear bracket  \cite{Sklyanin89}
\begin{equation}
\label{rLL}
\left[ L _1(\lambda) , L _2 (\mu) \right] = \left[ r_{12}(\lambda - \mu) , L _1(\lambda) + L _2 (\mu) \right].
\end{equation}

Using the expansion \eqref{expan-T} it is evident that
\begin{equation}
\label{expan-t}
t(\lambda) = 2 + \eta ^2  \sum _{m=1}^N \sum _{n \neq m}^N \frac{\vec{S}_{m} \cdot \vec{S}_{n}}{(\lambda - \alpha _m)(\lambda - \alpha _n)} + \mathcal{O}(\eta ^3) .
\end{equation}
The same expansion  \eqref{expan-T}  leads to
\begin{align}
\label{aux- expan-D}
\Delta \left[T(\lambda) \right]  &=   \mathbbm{1} + \eta\, \mathrm{tr} L(\lambda) + \frac{\eta ^2}{2} \mathrm{tr}_{00'} P_{00'}^- \sum _{m=1}^N  \left(  \frac{\vec{\sigma}_{0} \cdot \vec{S}_{m}}{(\lambda - \alpha _m)^2} -  \frac{ \vec{\sigma}_{0'} \cdot \vec{S}_{m}}{(\lambda - \alpha _m)^2} \right) \notag \\
&+ \frac{\eta ^2}{2} \mathrm{tr}_{00'} P_{00'}^- \sum _{m=1}^N \sum _{n \neq m}^N \left( \frac{\mathbbm{1}_0 \left( \vec{S}_{m} \cdot \vec{S}_{n} \right)}{(\lambda - \alpha _m)(\lambda - \alpha _n)} 
+ \frac{\mathbbm{1}_{0'} \left( \vec{S}_{m} \cdot \vec{S}_{n} \right)}{(\lambda - \alpha _m)(\lambda - \alpha _n)} \right) \notag \\ 
&+ \frac{\eta ^2}{2} \mathrm{tr}_{00'} P_{00'}^-  \sum _{m=1}^N \left(
\sum _{n > m}^N  \frac{\imath \vec{\sigma}_{0} \cdot \left( \vec{S}_{n} \times \vec{S}_{m} \right)}{(\lambda - \alpha _m)(\lambda - \alpha _n)} 
+  \sum _ {n<m}^N  \frac{\imath \vec{\sigma}_{0} \cdot \left( \vec{S}_{m} \times \vec{S}_{n} \right)}{(\lambda - \alpha _m)(\lambda - \alpha _n)} \right)  \notag \\ 
&+ \frac{\eta ^2}{2} \mathrm{tr}_{00'} P_{00'}^-  \sum _{m=1}^N \left(
\sum _{n > m}^N  \frac{\imath \vec{\sigma}_{0'} \cdot \left( \vec{S}_{n} \times \vec{S}_{m} \right)}{(\lambda - \alpha _m)(\lambda - \alpha _n)} 
+  \sum _ {n<m}^N  \frac{\imath \vec{\sigma}_{0'} \cdot \left( \vec{S}_{m} \times \vec{S}_{n} \right)}{(\lambda - \alpha _m)(\lambda - \alpha _n)} \right)  \notag \\ 
&+ {\eta ^2} \mathrm{tr}_{00'} P_{00'}^- L_0(\lambda)L_{0'}(\lambda) + \mathcal{O}(\eta ^3) ,
\end{align}
where $L(\lambda)$ is given in \eqref{L-Gaudin}. The final expression
for the expansion of $ \Delta \left[T(\lambda) \right]$ is obtained after taking all the traces 
\begin{equation}
\label{expan-D}
\Delta \left[T(\lambda) \right]  =   \mathbbm{1} +  \eta ^2 \left( \sum _{m=1}^N \sum _{n \neq m}^N \frac{\vec{S}_{m} \cdot \vec{S}_{n}}{(\lambda - \alpha _m)(\lambda - \alpha _n)} - \frac{1}{2} \mathrm{tr} \, L ^2 (\lambda) \right) + \mathcal{O}(\eta ^3). 
\end{equation}

To obtain the generation function of the Gaudin Hamiltonians notice that \eqref{expan-t} and \eqref{expan-D} yield
\begin{equation}
\label{exp-t-D}
t(\lambda) - \Delta \left[T(\lambda) \right] = \mathbbm{1} +  \frac{\eta ^2}{2} \mathrm{tr} \, L ^2 (\lambda) + \mathcal{O}(\eta ^3).
\end{equation}
Therefore   
\begin{equation}
\label{tau} 
\tau (\lambda) = \frac{1}{2} \mathrm{tr} \, L ^2 (\lambda)
\end{equation}
commute for different values of the spectral parameter,
\begin{equation}
\label{periodic-tau-tau}
[ \tau (\lambda) , \tau (\mu) ] = 0.
\end{equation}
Moreover, from \eqref{L-Gaudin} it is straightforward to obtain the expansion 
\begin{equation}
\label{poles-tau}
\tau (\lambda)  =  \sum _{m=1}^N \frac{2 H_{m}}{\lambda - \alpha _m} +  \sum _{m=1}^N \frac{\vec{S}_{m} \cdot \vec{S}_{m}}{(\lambda - \alpha _m)^2} =  \sum _{m=1}^N \frac{2 H_{m}}{\lambda - \alpha _m} 
+  \sum _{m=1}^N \frac{s_m (s_m+1)}{(\lambda - \alpha _m)^2}\,,
\end{equation}
and the Gaudin Hamiltonians, in the periodic case, are
\begin{equation}
\label{G-Ham}
H_m = \sum _{n \neq m}^N \frac{\vec{S}_{m} \cdot \vec{S}_{n}}{\alpha _m - \alpha _n}.
\end{equation}
This shows that $\tau (\lambda)$ is the generating function of Gaudin Hamiltonians when the periodic boundary conditions are imposed \cite{Sklyanin89}.

\section{Reflection equation}
A way to introduce non-periodic boundary conditions which are compatible with the integrability of the bulk model, was developed in \cite{Sklyanin88}. Boundary conditions on the left and right sites of the system are encoded in the left and right reflection matrices $K^-$ and $K^+$. The compatibility condition between the bulk and the boundary of the system takes the form of the so-called reflection equation. It is written in the following form for the left reflection matrix acting on the space $\mathbb{C}^2$ at the first site $K^-(\lambda) \in \mathrm{End} (\mathbb{C}^2)$
\begin{equation}
\label{RE}
R_{12}(\lambda - \mu) K^-_1(\lambda) R_{21}(\lambda + \mu) K^-_2(\mu)=
K^-_2(\mu) R_{12}(\lambda + \mu) K^-_1(\lambda) R_{21}(\lambda - \mu) .
\end{equation}

Due to the properties of the Yang R-matrix the dual reflection equation can be presented in the following form
\begin{equation}
\label{dRE}
R_{12}( \mu-\lambda )K_1^{+}(\lambda) R_{21}(-\lambda - \mu - 2\eta)  K_2^{+}(\mu)=
K_2^{+}(\mu) R_{12}(-\lambda -\mu-2\eta) K_1^{+}(\lambda) R_{21}(\mu-\lambda) .
\end{equation}
One can then verify that the mapping
\begin{equation}
\label{bijectionKpl}
K^+(\lambda)= K^{-}(- \lambda -\eta)
\end{equation}
is a bijection between solutions of the reflection equation and the dual reflection equation. After substitution of \eqref{bijectionKpl} into the dual reflection equation \eqref{dRE} one gets the reflection equation \eqref{RE} with shifted arguments.

The general, spectral parameter dependent solutions of the reflection equation \eqref{RE} can be written as follows \cite{VegaGonzalez}
\begin{equation}
\label{K-min}
K^{-}(\lambda) = \left(\begin{array}{cc}
\xi - \lambda & \psi \lambda \\ \phi \lambda & \xi + \lambda \end{array}\right) .
\end{equation}
It is straightforward to check the following useful identities
\begin{align}
\label{KKmin-l}
K^{-}(- \lambda)  K^{-}(\lambda) &= \left(\xi^2-\lambda ^2 \left( 1 + \phi \psi \right) \right) \mathbbm{1} 
= \mathrm{det} \left(K^{-}(\lambda) \right)  \mathbbm{1} , \\
\label{Kmin-l}    
K^{-}(- \lambda) &= \mathrm{tr} \, K^{-}(\lambda) - K^{-}(\lambda) .       
\end{align}

\section{$s\ell(2)$ Gaudin model with boundary terms}
With the aim of describing the inhomogeneous XXX spin chain with non-periodic boundary condition it is instructive to recall some properties of the Lax operator \eqref{L-XXX}. The identity \eqref{unit:Lax}
can be rewritten in the form \cite{CAMS}
\begin{equation}
\label{unit:LaxII}
\mathbb{L}_{0m}(\lambda - \alpha _m) \mathbb{L}_{0m}(-\lambda + \alpha _m + \eta) 
= \Big(1+\frac{\eta^2\,s_m (s_m+1)}{(\lambda - \alpha _m) (-\lambda + \alpha _m + \eta) } \Big) \mathbbm{1} _0 \, .
\end{equation}
It follows from the equation above and the RLL-relations \eqref{RLL} that the RTT-relations \eqref{RTT} can be recast as follows
\begin{align}
\label{tTRT}
\widetilde{T} _ {0'}(\mu ) R_{00'} ( \lambda + \mu) T _{0} (\lambda )  &= T _ {0}(\lambda ) R_{00'} ( \lambda + \mu) \widetilde{T} _ {0'}(\mu ) , \\
\label{tTtTR}
\widetilde{T} _{0} (\lambda ) \widetilde{T} _ {0'}(\mu ) R_{00'} (\mu - \lambda)   &= R_{00'} (\mu - \lambda) \widetilde{T} _ {0'}(\mu ) \widetilde{T} _ {0}(\lambda ) ,
\end{align}
where
\begin{equation}
\label{tilde-T}    
\widetilde{T}(\lambda ) = \mathbb{L}_{01}(\lambda + \alpha _1 + \eta) \cdots \mathbb{L}_{0N} (\lambda + \alpha _N + \eta) .
\end{equation}
The Sklyanin monodromy matrix $\mathcal{T}(\lambda)$ of the inhomogeneous XXX spin chain with non-periodic boundary consists of the two matrices $T(\lambda)$ \eqref{monodromy-T} and $\widetilde{T} _{0} (\lambda )$ \eqref{tilde-T} and a reflection matrix $K ^{-}(\lambda)$ \eqref{K-min},
\begin{equation}
\label{calT}
\mathcal{T}_0(\lambda) = T_0(\lambda) K _0^{-}(\lambda) \widetilde T_0(\lambda) .
\end{equation}
Using the RTT-relations \eqref{RTT}, \eqref{tTRT}, \eqref{tTtTR} and the reflection equation \eqref{RE} it is straightforward to show that the exchange relations of the monodromy matrix $\mathcal{T}(\lambda)$  in $V_0\otimes V_{0'}$ are \cite{CAMS}
\begin{equation}
\label{exchangeRE}
R _{00'}(\lambda - \mu) \mathcal{T}_{0} (\lambda) R _{0'0} (\lambda + \mu) \mathcal{T} _{0'} (\mu) = 
\mathcal{T}_{0'}(\mu) R _{00^{\prime}} (\lambda + \mu) \mathcal{T}_{0} (\lambda) 
R _{0'0} (\lambda - \mu) ,
\end{equation}
The open chain transfer matrix is given by the trace of $\mathcal{T}(\lambda)$ over the auxiliary space $V_0$ with an extra reflection matrix $K^+(\lambda)$ \cite{Sklyanin88},
\begin{equation}
\label{open-t}
t (\lambda) = \mathrm{tr}_0 \left( K^+(\lambda) \mathcal{T}(\lambda) \right).
\end{equation}
The reflection matrix $K^+(\lambda)$ \eqref{bijectionKpl} is the corresponding solution of the dual reflection equation \eqref{dRE}. The commutativity of the transfer matrix for different values of the spectral parameter
\begin{equation}
\label{open-tt}
[t (\lambda) , t (\mu)] = 0,
\end{equation}
is guaranteed by the dual reflection equation \eqref{dRE} and the exchange relations \eqref{exchangeRE} of the monodromy matrix $\mathcal{T}(\lambda)$.

The exchange relations \eqref{exchangeRE} admit a central element
\begin{equation}
\label{Delta-T-cal}
\Delta \left[\mathcal{T}(\lambda)\right] = \mathrm{tr}_{00'} P^{-}_{00'} \mathcal{T}_{0}(\lambda-\eta/2) R_{00'} (2\lambda) \mathcal{T}_{0'}(\lambda+\eta/2). 
\end{equation}

For the study of the open Gaudin model we impose
\begin{equation}
\label{normalizationKpl}
\lim_{\eta \to 0}\Big(  K^+(\lambda) K^{-} (\lambda)\Big)  = \left(\xi^2-\lambda ^2 \left( 1 + \phi \psi \right) \right) \mathbbm{1}.
\end{equation}
In particular, this implies that the parameters of the reflection matrices on the left and on the right end of the chain are the same. In general this not the case in  the study of the open spin chain. However, this condition is essential for the Gaudin model. Then we will write
\beq\label{K-min-notation}
K^-(\lambda)\equiv K(\lambda),
\eeq
so that 
\beq \label{def-Kplus}
K^+(\lambda)= K(-\lambda-\eta)= K(-\lambda) -\eta\, M
\mb{with} M= \left(\begin{array}{cc}
-1 & \psi  \\ \phi  &  1 \end{array}\right).
\eeq
Remark that the matrix $M$ obeys $M^2=(1+ \psi \phi) \mathbbm{1}$.

The expansion of $T(\lambda)$ is given in \eqref{expan-T}. It is easy to get the expansion for $\widetilde T(\lambda)$ as introduced in \eqref{tilde-T} and then, the one for $\mathcal{T}(\lambda)$. Using the relation \eqref{def-Kplus}, we deduce the expansion of $t (\lambda)$ \eqref{open-t} in powers of $\eta$:
\begin{align}
\label{expan-t-open}
t(\lambda) &= 2 \left(\xi^2 - \lambda ^2 \left( 1 + \phi \psi \right) \right) - 2 \eta \lambda  \left( 1 + \phi \psi \right) \notag \\
&- \eta ^2 \, \mathrm{tr}_0 \left( M_0 \left( L _0 (\lambda)  K_0(\lambda) -  K_0(\lambda) 
L _0 (-\lambda) \right) \right) \notag \\
&+ \eta ^2 \left(\xi^2-\lambda ^2 \left( 1 + \phi \psi \right) \right)
\underset {n \neq m} {\sum _{m,n=1}^N}  \left(
\frac{\vec{S}_{m} \cdot \vec{S}_{n}}{(\lambda - \alpha _m)(\lambda - \alpha _n)} 
+ \frac{\vec{S}_{m} \cdot \vec{S}_{n}}{(\lambda + \alpha _m)(\lambda + \alpha _n)} \right) \notag \\
&- \eta ^2 \mathrm{tr}_0 \, L _0 (\lambda) K_0(\lambda) L _0 (- \lambda) K_0(- \lambda) 
+ \mathcal{O}(\eta ^3). 
\end{align}

Our next step is to obtain the expansion of $\Delta \left[\mathcal{T}(\lambda)\right]$ \eqref{Delta-T-cal} in powers of $\eta$. We follow the analogous steps as for the periodic case, and after some tedious but straightforward calculations we get
\begin{align}
\label{expan-delta-open}
\Delta \left[\mathcal{T}(\lambda)\right] &= \lambda \left( \mathrm{tr}_0^2 K_0(\lambda) -  \mathrm{tr}_0 K _0^{2}(\lambda) \right) + 2\eta \lambda \mathrm{tr}_0  K_0(\lambda)  \ \mathrm{tr}_0 \left( L_0(\lambda) K_0(\lambda)  -  K_0(\lambda) L_0(- \lambda) \right) \notag \\
&-2\eta \lambda \Big( \mathrm{tr}_0 \left\{ L_0(\lambda) K _0^{2}(\lambda) \right\} - 
 \mathrm{tr}_0  \left\{  L_0(- \lambda) K _0^{2}(\lambda) \right\} \Big) 
 - \frac{\eta}{2} \mathrm{tr}_0  \, K _0^{2} (\lambda) 
\notag \\
&+ \eta^2  \lambda \underset {n \neq m} {\sum _{m,n=1}^N} \left( \frac{\vec{S}_{m} \cdot \vec{S}_{n}}{(\lambda - \alpha _m)(\lambda - \alpha _n)} + \frac{\vec{S}_{m} \cdot \vec{S}_{n}}{(\lambda + \alpha _m)(\lambda + \alpha _n)} \right) \mathrm{tr}_0 \, K_0(-\lambda) K_0(\lambda) 
\notag \\
&- 2 \eta^2 \lambda \, \mathrm{tr}_0 \, L _0 (\lambda) K_0(\lambda) L _0 (- \lambda) K_0(- \lambda) 
\notag \\
&-  \eta^2 \, \mathrm{tr}_0 \left\{ \big( L _0 (\lambda) K_0(\lambda) - K_0(\lambda)  L _0 (- \lambda) \big) K_0(\lambda)  \right\}
\notag \\
&+ \eta^2 \lambda \, \Big(\mathrm{tr}_0  \left\{ L _0 (\lambda) K_0(\lambda) - K_0(\lambda)  L _0 (- \lambda) \right\} \Big) ^2 
\notag \\
&- \eta^2 \lambda \, \mathrm{tr}_0  \left\{ \Big( L _0 (\lambda) K_0(\lambda) - K_0(\lambda)  L _0 (- \lambda) \Big) \Big( L _0 (\lambda) K_0(\lambda) - K_0(\lambda)  L _0 (- \lambda) \Big)\right\} \notag \\
&+ \frac{\eta^2 \lambda}{4}  \mathrm{tr}_0 \, M_0 ^2 + \mathcal{O}(\eta ^3).
\end{align}
Using the relations \eqref{KKmin-l} and \eqref{Kmin-l} the first term of the expansion above simplifies and the second and third term together turn out  to be propositional to the trace of $L(\lambda)$ \eqref{L-Gaudin} and therefore vanish, 

\begin{align}
\label{final-expan-delta-open}
\Delta \left[\mathcal{T}(\lambda)\right] &= 2 \lambda \left(\xi^2 - \lambda ^2 \left( 1 + \phi \psi \right) \right)
- \eta \left(\xi^2 + \lambda ^2 \left( 1 + \phi \psi \right) \right) \notag \\
&+ 2\eta^2 \lambda \left(\xi^2 - \lambda ^2 \left( 1 + \phi \psi \right) \right)
\underset {n \neq m} {\sum _{m,n=1}^N} \left( \frac{\vec{S}_{m} \cdot \vec{S}_{n}}{(\lambda - \alpha _m)(\lambda - \alpha _n)} + \frac{\vec{S}_{m} \cdot \vec{S}_{n}}{(\lambda + \alpha _m)(\lambda + \alpha _n)} \right)  \notag \\
&- 2 \eta^2 \lambda \, \mathrm{tr}_0 \, L _0 (\lambda) K_0(\lambda) L _0 (- \lambda) K_0(- \lambda) 
\notag \\
&-  \eta^2 \, \mathrm{tr}_0 \left( \left( L _0 (\lambda) K_0(\lambda) - K_0(\lambda)  L _0 (- \lambda) \right) K_0(\lambda)  \right)
\notag \\
&+ \eta^2 \lambda \, \mathrm{tr}_0  \left\{  \Big(\mathrm{tr}_{0'}  \left\{ L _{0'} (\lambda) K_{0'} (\lambda) 
- K_{0'}(\lambda)  L _{0'} (- \lambda) \right\} -  L _0 (\lambda) K_0(\lambda) + K_0(\lambda)  L _0 (- \lambda) \Big) \times \right. 
\notag \\
&\qquad\qquad\quad\times \left. \Big( L _0 (\lambda) K_0(\lambda) - K_0(\lambda)  L _0 (- \lambda) \Big)   \right\} 
\notag \\
&+ \frac{\eta^2 \lambda}{2} \left( 1 + \phi \psi \right) + \mathcal{O}(\eta ^3). 
\end{align} 

In order to simplify some formulae we introduce the following notation
\begin{equation}
\label{cal-L}
\mathcal{L} _0 (\lambda) = L _0 (\lambda) - K_0 (\lambda)  L _0 (- \lambda) K_0^{-1}(\lambda),\\
\end{equation}

Using the formulas \eqref{expan-t-open} and \eqref{final-expan-delta-open} we calculate the expansion in powers of $\eta$ of the difference
\begin{align}
\label{exp-t-D-op}
&2\lambda t(\lambda) - \Delta \left[\mathcal{T}(\lambda)\right]   = 2 \lambda \left(\xi^2 - \lambda ^2 \left( 1 + \phi \psi \right) \right) + \eta \left(\xi^2 - 3 \lambda ^2 \left( 1 + \phi \psi \right) \right) -2 \eta ^2 \lambda \, \mathrm{tr}_0 \left( M_0 \mathcal{L}_0 (\lambda) K_0 (\lambda) \right) \notag\\
&+ \eta ^2 \mathrm{tr}_0 \left( \mathcal{L}_0 (\lambda) K_0^{2} (\lambda) \right) - \eta ^2 \lambda \,\mathrm{tr}_0 \left( \left(  \mathrm{tr}_{0'} \left(  \mathcal{L}_{0'} (\lambda) K_{0'} (\lambda) \right) \mathbbm{1}_0 -  \mathcal{L}_0 (\lambda) K_0 (\lambda)  \right) \mathcal{L}_0 (\lambda) K_0(\lambda) \right)
 \notag\\
&- \frac{\eta ^2 \lambda}{2} \left( 1 + \phi \psi \right) + \mathcal{O}(\eta ^3) .
\end{align}

Actually the third and the fourth term in the expression above vanish
\begin{align}
\label{simplification}
&\mathrm{tr}_0 \left( \mathcal{L}_0 (\lambda) K_0^{2} (\lambda) \right) - 2 \lambda \, \mathrm{tr}_0 \left( M_0 \mathcal{L}_0 (\lambda) K_0(\lambda) \right) =   \mathrm{tr}_0 \left( \left( \mathcal{L}_0 (\lambda) K_0 (\lambda) \right) \left( K_0 (\lambda) - 2\lambda M _0 \right) \right) 
    \notag\\
    &=  \mathrm{tr}_0 \left( \mathcal{L}_0 (\lambda) K_0 (\lambda) K_0 (-\lambda) \right) 
    = \left(\xi^2 - \lambda ^2 \left( 1 + \phi \psi \right) \right)  \mathrm{tr}_0 \mathcal{L}_0 = 0,	
\end{align}
due to the fact that the $ \mathrm{tr}_0 \mathcal{L}_0$ is equal to zero. Therefore the expansion \eqref{exp-t-D-op} reads
\begin{align}
\label{exp-t-D-open}
2 \lambda t(\lambda) - \Delta \left[\mathcal{T}(\lambda)\right]   &= 2 \lambda \left(\xi^2 - \lambda ^2 \left( 1 + \phi \psi \right) \right) + \eta \left(\xi^2 - 3 \lambda ^2 \left( 1 + \phi \psi \right) \right)  \notag\\
&- \eta ^2 \lambda \, \mathrm{tr}_0 \left( \left(  \mathrm{tr}_{0'} \left(  \mathcal{L}_{0'} (\lambda) K_{0'} (\lambda) \right) \mathbbm{1}_0 -  \mathcal{L}_0 (\lambda) K_0 (\lambda)  \right) \mathcal{L}_0 (\lambda) K_0 (\lambda) \right) \notag\\
&- \frac{\eta ^2 \lambda}{2} \left( 1 + \phi \psi \right) + \mathcal{O}(\eta ^3) .
\end{align}
It is important to notice that using the following identity
\begin{equation}
\label{mat-id}
\mathrm{tr}_{0'} \left(  \mathcal{L}_{0'} (\lambda) K_{0'} (\lambda) \right) \mathbbm{1}_0 -  \mathcal{L}_0 (\lambda) K_0 (\lambda) = - K_0 (- \lambda) \mathcal{L}_0 (\lambda) ,
\end{equation}
the third term in \eqref{exp-t-D-open} can be simplified 
\begin{equation}
\label{ 3-rd-term-sim}
\mathrm{tr}_0 \, K_0 (- \lambda) \mathcal{L}_0 (\lambda) \mathcal{L}_0 (\lambda) K_0 (\lambda) =  \left( \xi^2 - \lambda ^2 \left( 1 + \phi \psi \right) \right)  \mathrm{tr}_0 \, \mathcal{L}_0 ^2(\lambda) .
\end{equation}
Finally, the expansion \eqref{exp-t-D-open} reads
\begin{align}
\label{final-exp-t-D-open}
2 \lambda t(\lambda) - \Delta \left[\mathcal{T}(\lambda)\right]   &= 2 \lambda \left(\xi^2 - \lambda ^2 \left( 1 + \phi \psi \right) \right) + \eta \left(\xi^2 - 3 \lambda ^2 \left( 1 + \phi \psi \right) \right)  \notag\\
&+ \eta ^2  \lambda \left( \xi^2 - \lambda ^2 \left( 1 + \phi \psi \right) \right)  \mathrm{tr}_0 \, \mathcal{L}_0 ^2(\lambda)  \notag\\
&- \frac{\eta ^2 \lambda}{2} \left( 1 + \phi \psi \right) + \mathcal{O}(\eta ^3) .
\end{align}
This shows that
\begin{equation}
\label{open-tau} 
\tau (\lambda) =  \mathrm{tr}_0 \, \mathcal{L}_0 ^2(\lambda) 
\end{equation}
commute for different values of the spectral parameter,
\begin{equation}
\label{open-tau-tau}
[ \tau (\lambda) , \tau (\mu) ] = 0.
\end{equation}
and therefore can be considered to be the generating function of Gaudin Hamiltonians with boundary terms. The multiplicative factor in \eqref{open-tau}, which is equal to the determinant of $K (\lambda)$, will be useful in the partial fraction decomposition of the generating function.

With the aim of obtaining the Gaudin Hamiltonians with the boundary terms from the generating function  \eqref{open-tau}, it is instructive to study the representation of $\mathcal{L}_0 (\lambda)$ \eqref{cal-L} in terms of the local spin operators
\begin{equation}
\label{cal-L-loc}
 \mathcal{L}_0 (\lambda) = \sum _{m=1}^N \left( \frac{\vec{\sigma}_{0} \cdot \vec{S}_{m}}{\lambda - \alpha _m}  +    \frac{\left(K _0 (\lambda) \vec{\sigma}_{0} K _0^{-1}(\lambda) \right) \cdot \vec{S}_{m}}{\lambda + \alpha _m}   \right) ,
\end{equation}
noticing that
\begin{equation}
\label{cal-L-local}
 \mathcal{L}_0 (\lambda) = \sum _{m=1}^N \left( \frac{\vec{\sigma}_{0} \cdot \vec{S}_{m}}{\lambda - \alpha _m}  +    \frac{ \vec{\sigma}_{0} \cdot  \left( K _m^{-1}(\lambda) \vec{S}_{m} K _m(\lambda)\right)}{\lambda + \alpha _m}   \right).
\end{equation}
Now it is straightforward to obtain the expression for the generating function \eqref{open-tau} in terms of the local operators
\begin{align}
\label{op-tau-lo}
\tau (\lambda)  &= 2 \sum _{m,n=1}^N \left( \frac{\vec{S}_{m} \cdot \vec{S}_{n}}{(\lambda - \alpha _m)(\lambda - \alpha _n)}   \right.
+  \frac{ \vec{S}_{m} \cdot  \left( K _n^{-1}(\lambda) \vec{S}_{n} K _n(\lambda) \right) +  
 \left( K _n^{-1}(\lambda) \vec{S}_{n} K _n(\lambda) \right)  \cdot  \vec{S}_{m}}{(\lambda - \alpha _m) (\lambda + \alpha _n)}   
\notag \\
&\qquad\qquad\quad + \left. \frac{\left( K _m^{-1}(\lambda) \vec{S}_{m} K _m(\lambda) \right) \cdot \left( K _n^{-1}(\lambda) \vec{S}_{n} K _n(\lambda) \right) }{(\lambda + \alpha _m) (\lambda + \alpha _n)} \right) .
\end{align}
It is important to notice that  \eqref{op-tau-lo} simplifies further
\begin{align}
\label{open-tau-loc}
\tau (\lambda)  &= 2 \sum _{m,n=1}^N \left( \frac{\vec{S}_{m} \cdot \vec{S}_{n}}{(\lambda - \alpha _m)(\lambda - \alpha _n)} +
\frac{\vec{S}_{m} \cdot \vec{S}_{n}}{(\lambda + \alpha _m) (\lambda + \alpha _n)} \right.
\notag \\
&\qquad\qquad\quad+  \left.  \frac{ \vec{S}_{m} \cdot  \left( K _n^{-1}(\lambda) \vec{S}_{n} K _n(\lambda) \right) +   \left( K _n^{-1}(\lambda) \vec{S}_{n} K _n(\lambda) \right)  \cdot  \vec{S}_{m}}{(\lambda - \alpha _m) (\lambda + \alpha _n)}    \right) .
\end{align}

The Gaudin Hamiltonians with the boundary terms are obtained from
 the residues of the generating function \eqref{open-tau-loc} at  poles $\lambda = \pm\alpha_m$ :
\begin{equation}
\label{res-Ham}
\mbox{Res}_{\lambda = \alpha_m} \tau (\lambda) \ =\  4 \, H_m
\mb{and}
\mbox{Res}_{\lambda = -\alpha_m} \tau (\lambda) \ =\ (-4) \, \widetilde H_m
\end{equation}
where
\begin{equation}
\label{op-Ham-a}
H_m = \sum _{n \neq m}^N \frac{\vec{S}_{m} \cdot \vec{S}_{n}}{\alpha _m - \alpha _n} + 
\sum _{n = 1}^N  \frac{ \vec{S}_{m} \cdot  \left( K _n^{-1}(\alpha_m) \vec{S}_{n} K _n(\alpha_m) \right) +   \left( K _n^{-1}(\alpha_m) \vec{S}_{n} K _n(\alpha_m) \right)  \cdot  \vec{S}_{m}}{2(\alpha_m + \alpha _n)} ,
\end{equation}
and  
\begin{equation}
\label{op-Ham-b}
\widetilde{H}_m = \sum _{n \neq m}^N \frac{\vec{S}_{m} \cdot \vec{S}_{n}}{\alpha _m - \alpha _n} + 
\sum _{n = 1}^N  \frac{ \vec{S}_{m} \cdot  \left( K _n^{-1}(-\alpha_m) \vec{S}_{n} K _n(-\alpha_m) \right) +   \left( K _n^{-1}(-\alpha_m) \vec{S}_{n} K _n(-\alpha_m) \right)  \cdot  \vec{S}_{m}}{2(\alpha_m + \alpha _n)} .
\end{equation}
The above Hamiltonians can be expressed in somewhat a more symmetric form
\begin{equation}
\label{open-Ham-a}
H_m = \sum _{n \neq m}^N \frac{\vec{S}_{m} \cdot \vec{S}_{n}}{\alpha _m - \alpha _n} + 
\sum _{n = 1}^N  \frac{ \left( K _m (\alpha_m) \vec{S}_{m} K _m^{-1}  (\alpha_m) \right)  \cdot  \vec{S}_{n} + \vec{S}_{n} \cdot  \left( K _m (\alpha_m) \vec{S}_{m} K _m^{-1}  (\alpha_m) \right) }{2(\alpha_m + \alpha _n)} ,
\end{equation}
and 
\begin{equation}
\label{open-Ham-b}
\widetilde{H}_m = \sum _{n \neq m}^N \frac{\vec{S}_{m} \cdot \vec{S}_{n}}{\alpha _m - \alpha _n} +  
\sum _{n = 1}^N  \frac{ \left( K _m (-\alpha_m) \vec{S}_{m} K _m^{-1} (-\alpha_m) \right)  \cdot  \vec{S}_{n} + \vec{S}_{n} \cdot  \left( K _m (-\alpha_m) \vec{S}_{m} K _m^{-1} (-\alpha_m) \right) }{2(\alpha_m + \alpha _n)} .
\end{equation}

The next step is to study the quasi-classical limit of the exchange relations \eqref{exchangeRE} with the aim of deriving relevant algebraic structure for the Lax operator \eqref{cal-L}.

\section{Linear bracket relations}

The implementation of the algebraic Bethe ansatz requires the commutation relations between the entries of the Lax operator \eqref{cal-L}. Our aim is to derive these relations as the quasi-classical limit of \eqref{exchangeRE}. As the first step in this direction we observe that using \eqref{classical-r} the reflection equation \eqref{RE} can be expressed as 
\begin{equation}
\label{RE-limit}
\begin{split}
\big( \mathbbm{1} - \eta r_{12}(\lambda - \mu) \big) K _1(\lambda) \big( \mathbbm{1} - \eta r_{21}(\lambda + \mu) \big) K_2(\mu) =\\
= K_2(\mu) \big( \mathbbm{1} - \eta r_{12}(\lambda + \mu) \big) K _1(\lambda) \big( \mathbbm{1} - \eta r_{21}(\lambda - \mu) \big)
\end{split}
\end{equation}
The conditions obtained from the zero and first order in $\eta$ are identically satisfied for the matrix $K (\lambda)$. 
In particular, it obeys the classical reflection equation \cite{Sklyanin86,Sklyanin87}:
\begin{equation}
\label{classical-RE}
\begin{split}
r_{12}(\lambda - \mu) K _1(\lambda) K_2(\mu) + K _1(\lambda) r_{21}(\lambda + \mu) K_2(\mu) = \\
= K_2(\mu) r_{12}(\lambda + \mu) K _1(\lambda) +  K_2(\mu) K _1(\lambda) r_{21}(\lambda - \mu) .
\end{split}
\end{equation}
The terms of the second order in $\eta$ in \eqref{RE-limit} are
\begin{equation}
\label{RE-2-nd-eta}
r_{12}(\lambda - \mu) K _1(\lambda) r_{21}(\lambda + \mu) K_2(\mu) =
K_2(\mu) r_{12}(\lambda + \mu) K _1(\lambda) r_{21}(\lambda - \mu)  .
\end{equation}
This equation is also satisfied by the the K-matrix \eqref{K-min} and the classical r-matrix \eqref{classical-r}. In addition, the classical r-matrix \eqref{classical-r} has the unitarity property 
\begin{equation}
\label{r-unitarity}
r_{21}(-\lambda) = - r_{12}(\lambda) ,
\end{equation}
and satisfies the classical Yang-Baxter equation
\begin{equation}
\label{classicalYBE}
[r_{13} (\lambda), r_{23}(\mu) ] + [r_{12}(\lambda - \mu), r_{13} (\lambda) +  r_{23}(\mu)] =0.
\end{equation}
Now we can proceed  to the derivation of the relevant linear bracket relations of the Lax operator \eqref{cal-L}. 
The desired relations can be obtained by writing the equation \eqref{exchangeRE} in the following form, using \eqref{classical-r}, 
\begin{equation}
\label{exchRE-limit}
\begin{split}
\left( \mathbbm{1} - \eta r_{00'}(\lambda - \mu) \right) \mathcal{T} _0(\lambda) \left( \mathbbm{1} - \eta r_{0'0}(\lambda + \mu) \right) \mathcal{T} _{0'}(\mu) =\\
= \mathcal{T} _{0'}(\mu) \left( \mathbbm{1} - \eta r_{00'}(\lambda + \mu) \right) \mathcal{T} _0(\lambda) \left( \mathbbm{1} - \eta r_{0'0}(\lambda - \mu) \right)
\end{split}
\end{equation}
and substituting the expansion of $\mathcal{T}(\lambda)$ \eqref{calT} in powers of $\eta$
\begin{equation}
\label{exc-cal-T}
\mathcal{T} (\lambda) = K (\lambda) + \eta \, \mathcal{L} (\lambda) K (\lambda) +  \frac{\eta^2}{2} \, \frac{d ^2\mathcal{T} (\lambda)}{d\eta ^2}  |_{\eta = 0} + \mathcal{O}(\eta ^3).
\end{equation}
The zero and first orders in $\eta$ are identically satisfied for the matrix $K (\lambda)$ defined in \eqref{K-min}. 
The relations we seek follow from the terms of the second order in $\eta$ in \eqref{exchRE-limit}. When the terms containing the second order derivatives of $\mathcal{T}$ are eliminated and the equation \eqref{RE-2-nd-eta} is used to eliminate the other two terms, there are ten terms remaining. Using twice the classical reflection equation \eqref{classical-RE} and the unitarity property \eqref{r-unitarity} one obtains 
\begin{align}
\label{rLL-1}
&\left(  \mathcal{L} _{0} (\lambda)  \mathcal{L} _{0'}(\mu) -  \mathcal{L} _{0'}(\mu) \mathcal{L} _{0}(\lambda) \right) K_{0} (\lambda) K_{0'} (\mu) = \left( r_{00'} (\lambda - \mu) \mathcal{L} _{0} (\lambda) - \mathcal{L} _{0} (\lambda)  r_{00'} (\lambda - \mu) \right) \times  
\notag \\
&\times K_{0} (\lambda) K_{0'} (\mu) + \left( \mathcal{L} _{0} (\lambda) K_{0'} (\mu) r_{00'} (\lambda + \mu) - K_{0'} (\mu) r_{00'} (\lambda + \mu)  \mathcal{L} _{0} (\lambda) \right) K_{0} (\lambda) 
\notag \\
&- \left( r_{0'0} (\mu - \lambda) \mathcal{L} _{0'} (\mu) - \mathcal{L} _{0'} (\mu)   r_{0'0} (\mu - \lambda) \right) K_{0} (\lambda) K_{0'} (\mu) + \left( K_{0} (\lambda)  r_{0'0} (\mu + \lambda) \mathcal{L} _{0'} (\mu) \right.
\notag \\
&- \left. \mathcal{L} _{0'} (\mu) K_{0} (\lambda)  r_{0'0} (\mu + \lambda) \right)  K_{0'} (\mu).
\end{align}
Multiplying both sides of the equation \eqref{rLL-1} from the right  by $K_{0}^{-1} (\lambda) K_{0'}^{-1} (\mu)$, \eqref{rLL-1} can be express as
\begin{equation}
\label{rLL-2}
\begin{split}
\left[ \mathcal{L} _{0} (\lambda) , \mathcal{L} _{0'}(\mu) \right] = \left[ r_{00'} (\lambda - \mu) -
K_{0'} (\mu) r_{00'} (\lambda + \mu) K_{0'}^{-1} (\mu) , \mathcal{L} _{0} (\lambda) \right] \\
- \left[ r_{0'0} (\mu - \lambda)  - 
K_{0} (\lambda)  r_{0'0} (\mu + \lambda) K_{0}^{-1} (\lambda) , \mathcal{L} _{0'} (\mu) \right] .
\end{split}
\end{equation}
Defining
\begin{equation}
\label{rk-class}
r_{00'}^{K} (\lambda , \mu) = r_{00'} (\lambda - \mu) - K_{0'} (\mu) r_{00'} (\lambda + \mu) K_{0'}^{-1} (\mu),
\end{equation}
\eqref{rLL-2} can be written as
\begin{equation}
\label{rLL}
\left[ \mathcal{L} _{0} (\lambda) , \mathcal{L} _{0'}(\mu) \right] = \left[ r_{00'}^{K} (\lambda , \mu) , \mathcal{L} _{0} (\lambda) \right] - \left[ r_{0'0}^{K} (\mu , \lambda) , \mathcal{L} _{0'} (\mu) \right] .
\end{equation}

The commutator \eqref{rLL} is obviously anti-symmetric. It obeys the Jacobi identity because the $r$-matrix \eqref{rk-class} satisfies the classical YB equation
\begin{equation}
\label{G-classYBE}
[ r^K_{32}(\lambda_3,\lambda_2) , r^K_{13} (\lambda_1,\lambda_3) ] + [r^K_{12}(\lambda_1, \lambda_2), r^K_{13} (\lambda_1,\lambda_3) + r^K_{23}(\lambda_2,\lambda_3) ] 
=0. 
\end{equation}
The commutator \eqref{rLL} can also be recasted as an $(\fr,\fs)$ Maillet algebra \cite{Mail}. In the following we study the algebraic Bethe ansatz based on the linear bracket \eqref{rLL}. 

%
%
%
%
\section{Algebraic Bethe Ansatz}
Our preliminary step in the implementation of the algebraic Bethe ansatz for the open Gaudin model is to bring the boundary K-matrix to the upper, or lower, triangular form. As it was pointed out in \eqref{K-min}, the general form of the K-matrix \eqref{K-min-notation} is
\begin{equation}
\label{K-general}
\widetilde{K} (\lambda) =  \left(\begin{array}{cc}
\xi  - \lambda & \widetilde{\psi} \lambda \\ \widetilde{\phi} \lambda & \xi + \lambda \end{array} \right) .
\end{equation}
It is straightforward to check that the matrix
\begin{equation}
\label{U-mat}
U = \left(\begin{array}{cc}  - 1 - \nu & \widetilde{\phi} \\ \widetilde{\phi}  &  - 1 - \nu  \end{array}\right) ,
\end{equation}
with $\nu = \sqrt{1 + \widetilde{\phi} \, \widetilde{\psi} \ }$, which does not depend on the spectral parameter $\lambda$, brings the K-matrix to the upper triangular form by the similarity transformation
\begin{equation}
\label{K-triangular}
K (\lambda) = U^{-1}  \widetilde{K} (\lambda) U  =  \left( \begin{array}{cc}
\xi  - \lambda \nu & \lambda \psi  \\ 0 & \xi + \lambda \nu \end{array} \right) ,
\end{equation}
where $\psi = \widetilde{\phi} + \widetilde{\psi} $. Evidently, the inverse matrix is
\begin{equation}
\label{inv-K-triangular}
K ^{-1}(\lambda) = \frac{1}{\xi ^2 - \lambda ^2 \nu ^2}\left(
\begin{array}{cc}
\xi + \lambda \nu & - \lambda \psi \\
0 & \xi - \lambda \nu 
\end{array} \right) .
\end{equation}
Direct substitution of the formulas above into \eqref{cal-L-loc},
\begin{equation}
\label{cal-L-local}
\mathcal{L} _{0} (\lambda) = 
\left(\begin{array}{rr} H(\lambda) & F(\lambda) \\ E(\lambda)  & -H(\lambda)\end{array}\right)
= \sum _{m=1}^N \left( \frac{\vec{\sigma}_{0} \cdot \vec{S}_{m}}{\lambda - \alpha _m}  +    \frac{K _0 (\lambda) \vec{\sigma}_{0} K ^{-1}_{0} (\lambda) \cdot \vec{S}_{m}}{\lambda + \alpha _m}   \right) ,
\end{equation}
yields the following local realisation for the entries of the Lax matrix
\begin{align}
\label{gen-E}
E (\lambda) &=  \sum _{m=1}^N \left ( \frac{S ^+_m}{\lambda - \alpha _m} + \frac{(\xi + \lambda \nu ) S ^+_m}{(\xi - \lambda \nu )(\lambda + \alpha _m)}
\right) , \\[1ex]
\label{gen-F}
F (\lambda) &= \sum _{m=1}^N \left ( \frac{S ^-_m}{\lambda - \alpha _m} + \frac{ (\xi - \lambda \nu)^2 
S ^-_m -  \lambda^2 \psi ^2 S ^+_m - 2 \lambda \psi (\xi - \lambda \nu ) S ^3_m}{(\xi + \lambda \nu )(\xi - \lambda \nu )(\lambda + \alpha _m)} \right) , \\[1ex]
\label{gen-H} 
H (\lambda) &= \sum _{m=1}^N \left ( \frac{S ^3_m}{\lambda - \alpha _m} + \frac{\lambda \psi \,S ^+_m + (\xi - \lambda \nu) S ^3_m}{(\xi - \lambda \nu) (\lambda + \alpha _m)}
\right) .
\end{align}
The linear bracket \eqref{rLL} based on the r-matrix $r_{00'}^{K} (\lambda , \mu)$ \eqref{rk-class}, corresponding to \eqref{K-triangular}, \eqref{inv-K-triangular} and the classical r-matrix \eqref{classical-r}, defines the Lie algebra relevant for the open Gaudin model
\begin{align}
\left[ E (\lambda) , E (\mu) \right] &= 0 ,  \\
\left[ H (\lambda) , E (\mu) \right] &= \frac{2}{\lambda^2 - \mu^2} \left( \lambda \, E (\mu) - \frac{\xi - \lambda \nu}{\xi - \mu \nu} \mu \, E (\lambda) \right),  \\[1ex]
\label{com-EF}
\left[ E (\lambda) , F (\mu) \right] &=  \frac{2\psi \mu}{(\lambda + \mu) (\xi + \mu \nu)} \, E (\lambda) + \frac{4}{\lambda^2 - \mu^2}  \left( \frac{\xi - \mu \nu}{\xi - \lambda \nu} \lambda \, H (\mu) -  \frac{\xi + \lambda \nu}{\xi + \mu \nu} \mu \, H (\lambda) \right) ,  \\[1ex]
\left[ H (\lambda) , H (\mu) \right] &= \frac{- \psi}{\lambda + \mu} \left( \frac{\lambda}{\xi - \lambda \nu} \, E (\mu) - \frac{\mu}{\xi - \mu \nu} \, E (\lambda) \right), 
\end{align}
\begin{align}
\left[ H (\lambda) , F (\mu) \right] &= \frac{ \psi}{\lambda + \mu}  \left( \frac{2\lambda}{\xi - \lambda \nu} \, H (\mu) - \frac{\psi \mu ^2}{\xi^2 - \mu^2 \nu^2} \, E (\lambda) \right)  - \frac{2}{\lambda^2 - \mu^2} 
\left( \lambda \, F (\mu) -  \frac{\xi + \lambda \nu }{\xi + \mu \nu} \mu \, F (\lambda) \right) ,  \\[1ex]
\left[ F (\lambda) , F (\mu) \right] &= \frac{ 2 \psi}{\lambda + \mu}  \left( \frac{\lambda}{\xi + \lambda \nu} \, F (\mu) - \frac{\mu}{\xi + \mu \nu} \, F (\lambda) \right)  - \frac{2 \psi ^2}{\lambda + \mu} \left( \frac{\lambda ^2}{\xi ^2- \lambda ^2 \nu ^2} H (\mu) -  \frac{\mu ^2}{\xi ^2 - \mu ^2 \nu ^2} H (\lambda) \right) .
\end{align}
Our next step is to introduce the new generators $e(\lambda) , h(\lambda)$ and $f(\lambda)$ as the following linear combinations of the original generators
\begin{equation}
\label{new-basis}
e(\lambda) = \frac{-\xi + \lambda \nu}{\lambda} E(\lambda) 
, \quad h(\lambda) = \frac{1}{\lambda} \left( H(\lambda) - \frac{\psi\lambda}{2\xi} \, E(\lambda) \right) , \quad 
f(\lambda) = \frac{1}{\lambda} \left( (\xi + \lambda \nu) F(\lambda) +  \psi \lambda H (\lambda) \right) .
\end{equation}
The key observation is that in the new basis  we have
\begin{equation}
\label{0-comm}
\left[ e(\lambda) , e(\mu) \right] = \left[ h(\lambda) , h(\mu) \right] = \left[ f(\lambda) , f(\mu) \right] = 0 .
\end{equation}
Therefore there are only three nontrivial relations
\begin{align}
\label{h-e}
\left[ h(\lambda) , e(\mu) \right] &= \frac{2}{\lambda^2 - \mu^2} \left( e (\mu) - e (\lambda) \right) ,  \\[1ex]
\label{h-f}
\left[ h(\lambda) , f(\mu) \right]  &=  \frac{- 2}{\lambda^2 - \mu^2} 
\left( f (\mu) - f (\lambda) \right)  - \frac{2 \psi \nu}{(\lambda^2 - \mu^2) \xi } \left( \mu ^2 h (\mu) - \lambda ^2 h (\lambda) \right) \notag \\
&-  \frac{ \psi ^2}{(\lambda^2 - \mu^2) \xi^2} \left( \mu^2 e (\mu) - \lambda ^2 e(\lambda)\right)
,  \\[1ex]
\label{e-f}
\left[ e(\lambda) , f(\mu) \right]  &=  \frac{2 \psi \nu}{(\lambda^2 - \mu^2) \xi}
\left( \mu ^2 e (\mu) - \lambda ^2 e (\lambda) \right) - \frac{4}{\lambda^2 - \mu^2} \left( (\xi ^2 - \mu ^2 \nu ^2) h (\mu) - (\xi ^2 - \lambda ^2 \nu ^2) h (\lambda) \right) .
\end{align}

In the Hilbert space $\mathcal{H}$ \eqref{H-space},  in every $V_ m = \mathbb{C}^{2s+1}$ there exists a vector $\omega_m \in V_ m$ such that
\begin{equation}
\label{S-on-om}
S^3_m \omega _m = s_m \omega _m  \quad \text{and}  \quad S^+_m \omega _m = 0 .
\end{equation}
We define a vector $\Omega _+$ to be
\begin{equation}
\label{Omega+}
\Omega _+ = \omega _1 \otimes \cdots \otimes \omega _N \in \mathcal{H}.
\end{equation} 
From the definitions above, the formulas \eqref{gen-E} - \eqref{gen-H} and  \eqref{new-basis} it is straightforward to obtain the action of the generators $e(\lambda)$  and $h(\lambda)$ on the vector $\Omega _+$
\begin{equation}
\label{action-Om}
e(\lambda) \Omega _+ = 0, \quad \text{and} \quad  h(\lambda) \Omega _+ = \rho (\lambda) \Omega _+,
\end{equation}
with
\begin{equation}
\label{rho}
\rho (\lambda) = \frac{1}{\lambda} \sum _{m=1}^N \left( \frac{s_m}{\lambda - \alpha _m} +  \frac{s_m}{\lambda + \alpha _m} \right) =  \sum _{m=1}^N \frac{2 s_m}{\lambda ^2- \alpha ^2_m}.
\end{equation}

The generating function of the Gaudin Hamiltonians \eqref{open-tau} in terms of the entries of the Lax matrix is given by
\begin{equation}
\label{open-t} 
\tau (\lambda) =  \mathrm{tr}_0 \, \mathcal{L}_0 ^2(\lambda) = 2 H^2(\lambda) + 2 F (\lambda) E (\lambda) + \left[ E (\lambda) , F (\lambda) \right] .
\end{equation} 
From \eqref{com-EF} we have that the last term is
\begin{equation}
\left[ E (\lambda) , F (\lambda) \right] = 2 \frac{\xi ^2 + \lambda ^2 \nu ^2}{(\xi ^2 - \lambda ^2 \nu ^2) \lambda} H(\lambda) - 2 H^{\prime}(\lambda) + \frac{\psi}{\xi + \lambda \nu} E (\lambda) ,
\end{equation}
and therefore the final expression is
\begin{equation}
\label{open-ta} 
\tau (\lambda) = 2 \left( H^2(\lambda) + \frac{\xi ^2 + \lambda ^2 \nu ^2}{(\xi ^2 - \lambda ^2 \nu ^2) \lambda} H(\lambda) -  H^{\prime}(\lambda)  \right) + \left( 2 F (\lambda) + \frac{\psi}{\xi + \lambda \nu} \right) E (\lambda) .
\end{equation} 

Our aim is to implement the algebraic Bethe ansatz  based on the Lie algebra \eqref{0-comm} - \eqref{e-f}. To this end we need to obtain the expression for the generating function $\tau (\lambda)$ in terms of the generators $e(\lambda) , h(\lambda)$ and $f(\lambda)$. The first step is to invert the relations \eqref{new-basis}
\begin{align}
\label{new-b-inv-E}
E(\lambda) &= \frac{-\lambda}{\xi - \lambda\nu} \, e(\lambda) ,  \\[1ex]
\label{new-b-inv-H}
H(\lambda) &= \lambda \left( h(\lambda) - \frac{\psi \lambda}{2\xi (\xi - \lambda \nu)} \, e (\lambda) \right) ,  \\[1ex]
\label{new-b-inv-F}
F(\lambda) &= \frac{\lambda}{\xi + \lambda\nu} \left( f(\lambda) - \psi \lambda \,  h (\lambda) + \frac{\psi ^2 \lambda ^2}{2 \xi (\xi - \lambda \nu)} e(\lambda) \right) .
\end{align}

In particular, we have
\begin{align}
\label{H2}
H^2(\lambda) &= \lambda ^2 \left( h ^2 (\lambda) - \frac{\psi \lambda}{2 \xi (\xi - \lambda \nu)} \left( 2 h (\lambda) e (\lambda) - \left[  h (\lambda) , e (\lambda)\right] \right) + \frac{\psi ^2 \lambda ^2}{4 \xi ^2 (\xi - \lambda \nu) ^2}  e ^2(\lambda)  \right)  \notag \\[1ex]
&=  \lambda ^2 \left( h ^2 (\lambda)  - \frac{\psi \lambda}{2 \xi (\xi - \lambda \nu)} \left( 2 h (\lambda) e (\lambda) + \frac {e^{\prime} (\lambda)}{\lambda}   \right) + \frac{\psi ^2 \lambda ^2}{4 \xi ^2 (\xi - \lambda \nu) ^2} e ^2(\lambda) \right) .   
\end{align}
Substituting \eqref{new-b-inv-E} -- \eqref{H2} into \eqref{open-ta} we obtain the desired expression for the generating function
\begin{equation}
\label{open-tau} 
\begin{split}
\tau (\lambda) &= 2 \lambda ^2 \left( h ^2(\lambda) + \frac{2 \nu ^2}{\xi ^2 - \lambda ^2 \nu ^2} h (\lambda) - \frac{ h ^{\prime}(\lambda)}{\lambda} \right) \\[1ex]
&- \frac{2 \lambda ^2}{\xi ^2 - \lambda ^2 \nu ^2} \left( f (\lambda) + \frac{\psi \lambda ^2 \nu}{\xi} h (\lambda) + \frac{\psi ^2 \lambda ^2}{4 \xi ^2} e (\lambda) - \frac{\psi \nu}{\xi} \right) e (\lambda) .
\end{split}
\end{equation}

An important initial observation in the implementation of the algebraic Bethe ansatz is that the vector  $\Omega _+$ \eqref{Omega+} is an eigenvector of the generating function $\tau (\lambda) $, to show this we use the expression above,  \eqref{action-Om}  and \eqref{rho}  
\begin{equation}
\label{tau-Om}
\tau (\lambda) \Omega _+ = \chi _0 (\lambda) \Omega _+ = 2 \lambda ^2 \left( \rho ^2 (\lambda) + \frac{2 \nu ^2\, \rho (\lambda)}{\xi ^2 - \lambda ^2 \nu ^2}   - \frac{\rho^{\prime} (\lambda)}{\lambda} \right) \Omega _+ ,
\end{equation}
using \eqref{rho} the eigenvalue $\chi _0 (\lambda)$ can be expressed as
\begin{equation}
\label{chi-0}
\chi _0 (\lambda) = 8 \lambda ^2 \left( \sum _{m=1}^N \frac{s_m (s_m+1)}{(\lambda ^2 - \alpha_m^2)^2} + \sum _{m=1}^N \frac{s_m}{\lambda ^2 - \alpha_m^2} \left( \frac{\nu ^2}{\xi ^2 - \lambda ^2 \nu ^2} +
\sum _{n\neq m}^N \frac{2s_n}{\alpha_m^2- \alpha_n^2} \right) \right) .
\end{equation}

An essential step in the algebraic Bethe ansatz is a definition of the corresponding Bethe vectors. In this case, they are symmetric functions of their arguments and are such that the off shell action of the generating function of the Gaudin Hamiltonians is as simple as possible. With this aim we attempt to show that the Bethe vector $\varphi _1(\mu)$ has the form
\begin{equation}
\label{phi-1}
\varphi _1(\mu) = \left( f(\mu) + c_1(\mu) \right) \Omega _+,
\end{equation}
where $c_1(\mu)$ is given by
\begin{equation}
\label{c-1}
c_1(\mu) = - \frac{\psi \nu}{\xi} \left( 1 - \mu^2 \rho(\mu) \right) .
\end{equation} 
Evidently, the action of the generating function of the Gaudin Hamiltonians reads
\begin{equation}
\label{tau-p1}
\tau (\lambda) \varphi _1(\mu) = \left[ \tau (\lambda) , f(\mu) \right] \Omega _+ + \chi _0 (\lambda) \varphi _1(\mu) .
 \end{equation}
A straightforward calculation show that  the commutator in the first term of \eqref{tau-p1} is given by
\begin{align}
\label{com-t-f}
\left[ \tau (\lambda) , f(\mu) \right] \Omega _+ &= - \frac{8 \lambda ^2}{\lambda ^2 - \mu ^2} \left( \rho (\lambda) + \frac{\nu ^2}{\xi ^2 - \lambda ^2 \nu ^2} \right) \varphi _1(\mu) \notag \\[1ex]
 &+ \frac{8 \lambda ^2(\xi ^2 - \mu ^2 \nu ^2)}{(\lambda ^2 - \mu ^2)(\xi ^2 - \lambda ^2 \nu ^2)} \left(  \rho (\mu) + \frac{\nu ^2}{\xi ^2 - \mu ^2 \nu ^2} \right) \varphi _1(\lambda) .
\end{align}
Therefore the action of the generating function $\tau (\lambda)$ on $\varphi _1(\mu)$ is given by
\begin{equation}
\label{tau-phi-1}
\tau (\lambda) \varphi _1(\mu) = \chi _1 (\lambda, \mu) \varphi _1(\mu) + \frac{8 \lambda ^2 (\xi ^2 - \mu ^2 \nu ^2)}{(\lambda ^2 - \mu ^2)(\xi ^2 - \lambda ^2 \nu ^2)} \left(  \rho (\mu) + \frac{\nu ^2}{\xi ^2 - \mu ^2 \nu ^2} \right) \varphi _1(\lambda), 
\end{equation}
with
\begin{equation}
\label{chi-1}
\chi _1 (\lambda, \mu) = \chi _0 (\lambda) - \frac{8 \lambda ^2}{\lambda ^2 - \mu ^2} \left( \rho (\lambda) + \frac{\nu ^2}{\xi ^2 - \lambda ^2 \nu ^2} \right).
\end{equation}
The unwanted term in \eqref{tau-phi-1} vanishes when the following Bethe equation is imposed on the parameter $\mu$,
\begin{equation}
\label{BE-1}
\rho (\mu) + \frac{\nu ^2}{\xi ^2 - \mu ^2 \nu ^2} = 0 .
\end{equation}
Thus we have shown that $\varphi _1(\mu)$ \eqref{phi-1} is the desired Bethe vector of the generating function $\tau (\lambda)$ corresponding to the eigenvalue $\chi _1 (\lambda, \mu)$.

We seek the Bethe vector $\varphi _2 (\mu _1, \mu _2)$ as the following symmetric function
\begin{equation}
\label{phi-2}
\varphi _2(\mu _1, \mu _2) = f(\mu _1) f(\mu _2) \Omega _+ 
+ c_2^{(1)}(\mu _2 ; \mu _1) f(\mu _1) \Omega _+
+ c_2^{(1)}(\mu _1 ; \mu _2) f(\mu _2) \Omega _+
+ c_2^{(2)}(\mu _1, \mu _2) \Omega _+,
\end{equation}
where the scalar coefficients $c_2^{(1)}(\mu _1 ; \mu _2)$  and $c_2^{(2)}(\mu _1, \mu _2)$ are
\begin{align}
\label{c2-1}
c_2^{(1)}(\mu _1 ; \mu _2)  &= - \frac{\psi \nu}{\xi} \left( 1 - \mu_1^2 \rho(\mu_1) + \frac{2\mu_1^2 }{\mu_1^2 - \mu_2^2} \right) , \\
\label{c2-2aux}
c_2^{(2)}(\mu _1, \mu _2)  &= - \frac{\psi ^2}{\nu ^2} \left( \frac{(\xi ^2 - 3 \mu_2 ^2 \nu ^2)\rho(\mu_1) 
- (\xi ^2 - 3 \mu_1 ^2 \nu ^2)\rho(\mu_2)}{\mu_1^2 - \mu_2^2}
+ (\xi ^2 - ( \mu_1 ^2 + \mu_2 ^2 ) \nu ^2) \rho(\mu_1) \rho(\mu_2) \right) .
\end{align}
One way to obtain the action of $\tau (\lambda)$ on $\varphi _2(\mu _1, \mu _2)$ is to write
\begin{equation}
\label{calc-tau-phi-2}
\begin{split}
\tau (\lambda) \varphi _2 (\mu _1, \mu _2) &= \left[ \left[ \tau (\lambda) , f(\mu _1) \right] , f(\mu _2) \right] \Omega _+  + \left( f(\mu _2) + c_2^{(1)}(\mu _2 ; \mu _1) \right) \left[ \tau (\lambda) , f(\mu _1) \right] \Omega _+ \\[1ex]
& + \left( f(\mu _1) + c_2^{(1)}(\mu _1 ; \mu _2) \right) \left[ \tau (\lambda) , f(\mu _2) \right] \Omega _+
+ \chi _0 (\lambda) \varphi _2 (\mu _1, \mu _2) .
\end{split}
\end{equation}
Then to substitute \eqref{com-t-f} in the second and third term above and use the relation
\begin{equation}
\label{rec-phi-2}
\begin{split}
\left( f(\mu _1) + c_2^{(1)}(\mu _1 ; \mu _2) \right)  \varphi _1 (\mu _2) &= \varphi _2 (\mu _1, \mu _2) -
 \frac{\psi \nu}{\xi} \frac{2 \mu_2^2}{\mu_1^2-\mu_2^2} \varphi _1 (\mu _1) \\[1ex]
 &- \left( c_2^{(2)}(\mu _1, \mu _2) - c_1(\mu _1) c_1(\mu _2) + 2 \frac{\psi \nu}{\xi} \frac{\mu_1^2c_1(\mu _2) - \mu_2^2c_1(\mu _1) }{\mu_1^2-\mu_2^2} \right) \Omega _+ ,
\end{split} 
\end{equation} 
which follows from the definition \eqref{phi-2}. A straightforward calculation shows that the off shell action of the generating function $\tau (\lambda)$ on $\varphi _2 (\mu _1, \mu _2)$ is given by
\begin{equation}
\label{tau-phi-2}
\begin{split}
\tau (\lambda) \varphi _2 (\mu _1, \mu _2) &= \chi _2 (\lambda, \mu _1, \mu _2 ) \varphi _2 (\mu _1, \mu _2) 
+ \sum _{i=1}^2 \frac{8 \lambda ^2 (\xi ^2 - \mu_i ^2 \nu ^2)}{(\lambda ^2 - \mu_i  ^2)(\xi ^2 - \lambda ^2 \nu ^2)} \times \\
& \times \left(  \rho (\mu_i ) + \frac{\nu ^2}{\xi ^2 - \mu_i ^2 \nu ^2} - \frac{2}{\mu_i ^2 - \mu _{3-i}^2}
\right) \varphi _2 (\lambda, \mu _{3-i}) , 
\end{split}
\end{equation}
with the eigenvalue
\begin{equation}
   \label{chi-2}
\chi _2 (\lambda, \mu _1, \mu _2) = \chi _0 (\lambda) - \sum _{i=1}^2 \frac{8 \lambda ^2}{\lambda ^2 - \mu_i ^2} \left( \rho (\lambda) + \frac{\nu ^2}{\xi ^2 - \lambda ^2 \nu ^2} - \frac{1}{\lambda ^2 - \mu _{3-i}^2} \right).   
\end{equation}
The two unwanted terms in the action above \eqref{tau-phi-2} vanish when the Bethe equations are imposed on the parameters $\mu _1$ and $\mu _2$,
\begin{equation}
\label{BE-2}
\rho (\mu _i) + \frac{\nu ^2}{\xi ^2 - \mu_i ^2 \nu ^2}  - \frac{2}{\mu_i ^2 - \mu _{3-i}^2} = 0 ,
\end{equation}
with $i=1,2$. Therefore $\varphi _2 (\mu _1, \mu _2)$ is the Bethe vector of the generating function of the Gaudin Hamiltonians with the eigenvalue $\chi _2 (\lambda, \mu _1, \mu _2)$.

As our next step we propose the Bethe vector $\varphi _3 (\mu _1, \mu _2, \mu _3)$ in the form of the following symmetric function of its arguments
\begin{equation}
\label{phi-3}
\begin{split}
\varphi _3 (\mu _1, \mu _2, \mu _3) &= f(\mu _1) f(\mu _2) f(\mu _3) \Omega _+ 
+ c_3^{(1)}( \mu _1 ; \mu _2 , \mu _3) f(\mu _2) f(\mu _3) \Omega _+
+ c_3^{(1)}( \mu _2 ; \mu _3 , \mu _1) f(\mu _3) f(\mu _1) \Omega _+ \\
&+ c_3^{(1)}( \mu _3 ; \mu _1 , \mu _2) f(\mu _1) f(\mu _2) \Omega _+
+ c_3^{(2)}(\mu _1 , \mu _2 ; \mu _3) f(\mu _3) \Omega _+
+ c_3^{(2)}(\mu _2 , \mu _3 ; \mu _1) f(\mu _1) \Omega _+ \\
&+ c_3^{(2)}(\mu _3 , \mu _1 ; \mu _2) f(\mu _2) \Omega _+
+ c_3^{(3)}(\mu _1, \mu _2,  \mu _3 ) \Omega _+,
\end{split}
\end{equation}
where the three scalar coefficients above are given by

\begin{align}
\label{c3-1}
c_3^{(1)}( \mu _1 ; \mu _2 , \mu _3) &= - \frac{\psi \nu}{\xi} \left( 1 - \mu_1^2 \rho(\mu_1) + \frac{2\mu_1^2 }{\mu_1^2 - \mu_2^2} + \frac{2\mu_1^2 }{\mu_1^2 - \mu_3^2}
\right) , \\[2ex]
\label{c3-2}
c_3^{(2)}(\mu _1 , \mu _2 ; \mu _3)  &= - \frac{\psi ^2}{\nu ^2} \left( \frac{\xi ^2 - 3 \mu_2 ^2 \nu ^2}{\mu_1^2 - \mu_2^2} \left( \rho (\mu _1) - \frac{2}{\mu_1 ^2 - \mu _3^2} \right) 
- \frac{\xi ^2 - 3 \mu_1 ^2 \nu ^2}{\mu_1^2 - \mu_2^2} \left( \rho (\mu _2) - \frac{2}{\mu_2 ^2 - \mu _3^2} \right) \right) 	\notag \\
&- \frac{\psi ^2}{\nu ^2} (\xi ^2 - ( \mu_1 ^2 + \mu_2 ^2 ) \nu ^2) \left( \rho (\mu _1) - \frac{2}{\mu_1 ^2 - \mu _3^2} \right) 
\left( \rho (\mu _2) - \frac{2}{\mu_2 ^2 - \mu _3^2} \right) ,
\end{align}
\begin{align}
\label{c3-3}
&c_3^{(3)}(\mu _1 , \mu _2 , \mu _3)  = - \frac{\psi^3}{\nu ^3 \xi}	 
\left( \frac{4 \xi ^4 +  \left( \xi ^2 + \mu _1^2 \nu ^2 \right) \left( 4 \mu _1^2 - 5 ( \mu _2^2 +  \mu _3^2) \right) \nu ^2}{(\mu _1 ^2 - \mu _2 ^2)(\mu _1 ^2 - \mu _3 ^2)} \rho (\mu _1) \right.
\notag \\
&\left. + \frac{4 \xi ^4 + \left( \xi ^2 + \mu _2^2 \nu ^2 \right) \left( 4 \mu _2^2 - 5 ( \mu _3^2 +  \mu _1^2) \right) \nu ^2}{(\mu _2 ^2 - \mu _3 ^2)(\mu _2 ^2 - \mu _1 ^2)} \rho (\mu _2) 
+ \frac{4 \xi ^4 + \left( \xi ^2 + \mu _3^2 \nu ^2 \right) \left( 4 \mu _3^2 - 5 ( \mu _1^2 +  \mu _2^2) \right) \nu ^2}{(\mu _3 ^2 - \mu _1 ^2)(\mu _3 ^2 - \mu _2 ^2)} \rho (\mu _3)
\right)	\notag \\
&- \frac{\psi^3}{\nu ^3 \xi}	\left( \frac{\xi ^2 \nu ^2 \left( \mu _1 ^4 + \mu _2 ^4 - \mu _1 ^2 \mu _2 ^2 + 2 \mu _3^2 \left( \mu _1 ^2 + \mu _2 ^2 \right) - 5 \mu _3^4 \right)
- \left( 2 \xi ^4  - \mu _1 ^2 \mu _2 ^2 \nu ^4 \right) \left( \mu _1 ^2 + \mu _2 ^2 - 2 \mu _3^2 \right)}
{(\mu _1 ^2 - \mu _3 ^2)(\mu _2 ^2 - \mu _3 ^2)} \rho (\mu _1) \rho (\mu _2 ) \right. \notag \\
&+ \frac{\xi ^2 \nu ^2 \left( \mu _2 ^4 + \mu _3 ^4 - \mu _2 ^2 \mu _3 ^2 + 2 \mu _1^2 \left( \mu _2 ^2 + \mu _3 ^2 \right) - 5 \mu _1^4 \right)
- \left( 2 \xi ^4  - \mu _2 ^2 \mu _3 ^2 \nu ^4 \right) \left( \mu _2 ^2 + \mu _3 ^2 - 2 \mu _1^2 \right)}
{(\mu _2 ^2 - \mu _1 ^2)(\mu _3 ^2 - \mu _1 ^2)}   \rho (\mu _2) \rho (\mu _3 ) \notag \\
&+\left. \frac{\xi ^2 \nu ^2 \left( \mu _3 ^4 + \mu _1 ^4 - \mu _3 ^2 \mu _1 ^2 + 2 \mu _2^2 \left( \mu _3 ^2 + \mu _1 ^2 \right) - 5 \mu _2^4 \right)
- \left( 2 \xi ^4  - \mu _3 ^2 \mu _1 ^2 \nu ^4 \right) \left( \mu _3 ^2 + \mu _1 ^2 - 2 \mu _2^2 \right)}
{(\mu _3 ^2 - \mu _2 ^2)(\mu _1 ^2 - \mu _2 ^2)}  \rho (\mu _3) \rho (\mu _1 ) \right) \notag \\
&- \frac{\psi^3}{\nu ^3} \xi \left(2 \xi ^2 - \left( \mu _1 ^2 + \mu _2 ^2 +  \mu _3 ^2 \right) \nu ^2 \right) \rho (\mu _1) \rho (\mu _2 ) \rho (\mu _3) .
\end{align}
A lengthy but straightforward calculation based on appropriate generalisation of \eqref{calc-tau-phi-2} and \eqref{rec-phi-2} shows that the action of the generating function $\tau (\lambda)$ on $\varphi _3 (\mu _1, \mu _2, \mu _3)$ is given by
\begin{equation}
\label{tau-phi-3}
\begin{split}
\tau (\lambda) \varphi _3 (\mu _1, \mu _2, \mu _3) &= \chi _3 (\lambda,\mu _1, \mu _2 \mu _3) 
\varphi _3 (\mu _1, \mu _2, \mu _3) + \sum _{i=1}^3 \frac{8 \lambda ^2 (\xi ^2 - \mu_i ^2 \nu ^2)}{(\lambda ^2 - \mu_i  ^2)(\xi ^2 - \lambda ^2 \nu ^2)} \times \\
& \times \left(  \rho (\mu_i ) + \frac{\nu ^2}{\xi ^2 - \mu_i ^2 \nu ^2} - \sum _{j\neq i}^3 \frac{2}{\mu_i ^2 - \mu _j^2}
\right) \varphi _3 (\lambda, \{ \mu _j \} _{j\neq i} ) , 
\end{split}
\end{equation}
where the eigenvalue is
\begin{equation}
   \label{chi-3}
\chi _3 (\lambda,\mu _1, \mu _2, \mu _3) = \chi _0 (\lambda) - \sum _{i=1}^3 \frac{8 \lambda ^2}{\lambda ^2 - \mu_i ^2} \left( \rho (\lambda) + \frac{\nu ^2}{\xi ^2 - \lambda ^2 \nu ^2} - \sum _{j\neq i}^3 \frac{1}{\lambda ^2 - \mu _j^2} \right).   
\end{equation}
The three unwanted terms in \eqref{tau-phi-3} vanish when the Bethe equation are imposed on the parameters $\mu _i$,
\begin{equation}
\label{BE-3}
\rho (\mu _i) + \frac{\nu ^2}{\xi ^2 - \mu_i ^2 \nu ^2}  -\sum _{j\neq i}^3 \frac{2}{\mu_i ^2 - \mu _j^2} = 0 ,
\end{equation}
with $i=1,2,3$. 

As a symmetric function of its arguments the Bethe vector $\varphi _4 (\mu _1, \mu _2, \mu _3, \mu _4)$ is given explicitly in the Appendix B. It is possible to check that the off shell action of the generating function $\tau (\lambda)$ on the Bethe vector $\varphi _4 (\mu _1, \mu _2, \mu _3, \mu _4)$ is given by
\begin{equation}
\label{tau-phi-4}
\begin{split}
\tau (\lambda) \varphi _4 (\mu _1, \mu _2, \mu _3, \mu _4) &= \chi _4 (\lambda,\mu _1, \mu _2 \mu _3) 
\varphi _4 (\mu _1, \mu _2, \mu _3, \mu _4) + \sum _{i=1}^4 \frac{8 \lambda ^2 (\xi ^2 - \mu_i ^2 \nu ^2)}{(\lambda ^2 - \mu_i  ^2)(\xi ^2 - \lambda ^2 \nu ^2)} \times \\
& \times \left(  \rho (\mu_i ) + \frac{\nu ^2}{\xi ^2 - \mu_i ^2 \nu ^2} - \sum _{j\neq i}^4 \frac{2}{\mu_i ^2 - \mu _j^2}
\right) \varphi _4 (\lambda, \{ \mu _j \} _{j\neq i} ) , 
\end{split}
\end{equation}
with the eigenvalue 
\begin{equation}
   \label{chi-4}
\chi _4 (\lambda,\mu _1, \mu _2 , \mu _3, \mu _4) = \chi _0 (\lambda) - \sum _{i=1}^4 \frac{8 \lambda ^2}{\lambda ^2 - \mu_i ^2} \left( \rho (\lambda) + \frac{\nu ^2}{\xi ^2 - \lambda ^2 \nu ^2} - \sum _{j\neq i}^4 \frac{1}{\lambda ^2 - \mu _j^2} \right). 
\end{equation}
The four unwanted terms on the right hand side of \eqref{tau-phi-4} vanish when the Bethe equation are imposed on the parameters $\mu _i$,
\begin{equation}
\label{BE-4}
\rho (\mu _i) + \frac{\nu ^2}{\xi ^2 - \mu_i ^2 \nu ^2}  -\sum _{j\neq i}^4 \frac{2}{\mu_i ^2 - \mu _j^2} = 0 ,
\end{equation}
with $i=1,2,3,4$. 

Based on the results presented above we can conclude that the local realisation \eqref{gen-E} - \eqref{gen-H} of the Lie algebra \eqref{new-basis} - \eqref{e-f} yields the spectrum $\chi _M (\lambda , \mu _1, \dots , \mu _M)$ of the generating function of the Gaudin Hamiltonians \begin{equation}
\label{chi-M}
\chi _M (\lambda , \mu _1, \dots , \mu _M) = \chi _0 (\lambda) - \sum _{i=1}^M \frac{8 \lambda ^2}{\lambda ^2 - \mu_i ^2} \left( \rho (\lambda) + \frac{\nu ^2}{\xi ^2 - \lambda ^2 \nu ^2} - \sum _{j\neq i}^M \frac{1}{\lambda ^2 - \mu _j^2} \right) ,
\end{equation}
and the corresponding Bethe equations which should be imposed on the parameters $\mu_i$
\begin{equation}
\rho (\mu _i) + \frac{\nu ^2}{\xi ^2 - \mu_i ^2 \nu ^2}  -\sum _{j\neq i}^M \frac{2}{\mu_i ^2 - \mu _j^2} = 0 ,
\end{equation}
where $i =1, 2, \dots , M$. Moreover, from \eqref{res-Ham} and \eqref{chi-M} it follows that the eigenvalues of the Gaudin Hamiltonians \eqref{open-Ham-a} and \eqref{open-Ham-b} can be obtained as the residues of $\chi _M (\lambda , \mu _1, \dots , \mu _M)$ at poles $\lambda = \pm \alpha _m$
\begin{equation}
\label{E-m}
\begin{split}
\mathcal{E} _m = \frac{1}{4} \mbox{Res}_{\lambda = \alpha_m} \chi _M (\lambda , \mu _1, \dots , \mu _M) &= \frac{s_m(s_m +1)}{2 \alpha _m} + \alpha _m s_m \left( \frac{\nu ^2}{\xi ^2 - \alpha _m ^2 \nu ^2}
+ \sum _{n\neq m} ^N \frac{2 s_n}{\alpha _m^2 -\alpha _n^2}  \right) \\ 
&- 2 \alpha _m s_m \sum _{i=1}^M \frac{1}{\alpha _m ^2 - \mu _i ^2} ,
\end{split}
\end{equation}
and
\begin{equation}
\label{tilde-E-m}
\begin{split}
\widetilde{\mathcal{E}} _m = - \frac{1}{4} \mbox{Res}_{\lambda = -\alpha_m} \chi _M (\lambda , \mu _1, \dots , \mu _M) &=  \frac{s_m(s_m +1)}{2 \alpha _m} + \alpha _m s_m \left(  \frac{\nu ^2}{\xi ^2 - \alpha _m ^2 \nu ^2} + \sum _{n\neq m} ^N \frac{2 s_n}{\alpha _m^2 -\alpha _n^2} \right) 
\\ 
&- 2 \alpha _m s_m \sum _{i=1}^M \frac{1}{\alpha _m ^2 - \mu _i ^2}.
\end{split}
\end{equation}
Evidently, the respective eigenvalues \eqref{E-m} and \eqref{tilde-E-m} of the Hamiltonians \eqref{open-Ham-a} and \eqref{open-Ham-b} coincide. When all the spin $s_m$ are set to one half, these energies coincide with the expressions obtained in \cite{HaoCaoYang14} (up to normalisation).
The Bethe equations are also equivalent, the correspondence between our variables and the one used in \cite{HaoCaoYang14} being given by (the left hand sides correspond to our variables, the left hand sides to the ones used in \cite{HaoCaoYang14}):
\begin{equation}
\mu_j= \frac{\lambda_j}{1-\xi^{(1)} } \ ; \quad \alpha_m= \frac{\theta_m}{1-\xi^{(1)}} \ ; \quad \frac{\xi}{\nu} = \frac{\xi}{1-\xi^{(1)}} .
\end{equation}

However, explicit and compact form of the relevant Bethe vector $\varphi _M (\mu _1, \mu _2, \dots , \mu _M)$, for an arbitrary positive integer $M$, requires further studies and will be reported elsewhere. As it is evident form the formulas for the Bethe vector $\varphi _4 (\mu _1, \mu _2, \mu _3, \mu _4)$ given in the Appendix B, the main problem lies in the definition the scalar coefficients $c_M^{(m)}(\mu _1 , \dots \mu _m ; \mu _{m+1}, \dots, \mu _M)$, with $m = 1, 2, \dots , M$. Some of them can be obtained straightforwardly, but, in particular, the coefficient   $c_M^{(M)}(\mu _1 , \mu _2 \dots , \mu _M)$ still represents a challenge, at least in the present form of the Bethe vectors.

\section{Conclusion}
Following Sklyanin's proposal in the periodic case \cite{Sklyanin89}, here we have derived the generating function of the Gaudin Hamiltonians with boundary terms. Our derivation is based on the quasi-classical expansion of the linear combination of the transfer matrix of the XXX Heisenberg spin chain and the central element, the so-called Sklyanin determinant. The corresponding Gaudin Hamiltonians with boundary terms are obtained as the residues of the generating function. Then we have studied the appropriate algebraic structure, including the classical reflection equation. Our approach to the algebraic Bethe ansatz is based on the relevant Lax matrix which satisfies certain
linear bracket and simultaneously provides the local realisation for the corresponding Lie algebra.
By defining the appropriate Bethe vectors we have obtained the strikingly simple off shell action of the generating function of the Gaudin Hamiltonians. Actually, the action of the generating function is as simple as it could possible be since it almost coincides with the one in the case when the boundary matrix is diagonal \cite{Hikami95}. In this way we have implemented the algebraic Bethe ansatz, obtaining the spectrum of the generating function and the corresponding Bethe equations.

Although the off shell action of the generating function which we have established is very simple, it would be important to obtain more compact formula for the Bethe vector $\varphi _M (\mu _1, \mu _2, \dots , \mu _M)$, for an arbitrary positive integer $M$. In particular,  simpler expression for the scalar coefficients $c_M^{(m)}(\mu _1 , \dots \mu _m ; \mu _{m+1}, \dots, \mu _M)$, with $m = 1, 2, \dots , M$ would be of utmost importance. Such a formula would be crucial for the off shell scalar product of the Bethe vectors and these results could lead to the correlations functions of Gaudin model 
with boundary. Moreover, it would be of considerable interest to establish a relation between Bethe vectors and solutions of the corresponding Knizhnik-Zamolodchikov equations, along the lines it was done in the case  when the boundary matrix is diagonal \cite{Hikami95}.

\bigskip

\noindent
\textbf{Acknowledgments}

\noindent
We acknowledge useful discussions with Zolt\'an Nagy. E.R. would like to thank to the staff of the GFM-UL and the Department of Mathematics of the University of the Algarve for warm hospitality while a part of the work was done. I. S. was supported in part by the Serbian Ministry of Science and Technological Development under grant number ON 171031.\break N. M. is thankful to Professor Victor Kac and the staff of the Mathematics Department at MIT for their kind hospitality. N. M. was supported in part by the FCT sabbatical fellowship SFRH/BSAB/1366/2013.


%
%
%
%
\appendix
\section{Basic definitions}
We consider the spin operators 
$S^{\alpha}$ 
with $\alpha = +, - , 3$, acting in some (spin $s$) representation space $\mathbb{C}^{2s+1}$ with the commutation relations
\begin{equation}
\label{crspin1}
[S^3, S^{\pm}] = \pm S^{\pm}, \quad [S^+,S^-] = 2 S^3 , 
\end{equation}
and Casimir operator
$$c_2 = (S^3) ^2 + \frac{1}{2} (S^+S^-+S^-S^+) = (S^3) ^2 + S^3 + S^-S^+=\vec{S}\cdot\vec{S}.$$
In the particular case of spin $\frac12$ representation, one recovers the Pauli matrices
$$
S^{\alpha} = \frac{1}{2} \sigma ^{\alpha} = \frac{1}{2} \left(\begin{array}{cc}
\delta_{\alpha3} & 2\delta_{\alpha+}  \\
2\delta_{\alpha-} & - \delta _{\alpha 3} \end{array}\right).
$$ 

We consider a spin chain with N sites with spin $s$ representations, i.e. a local $\mathbb{C}^{2s+1}$ space at each site and the operators 
\begin{equation}
S_m^{\alpha} = \mathbbm{1} \otimes \cdots \otimes \underbrace{S^{\alpha}} _m \otimes \cdots \otimes \mathbbm{1},
\end{equation}
with $\alpha = +,-, 3$ and $m= 1, 2 ,\dots , N$.

\section{Bethe vector $\varphi _4 (\mu _1, \mu _2, \mu _3, \mu _4)$}

Here we present explicit formulas of the Bethe vector $\varphi _4 (\mu _1, \mu _2, \mu _3, \mu _4)$. The vector $\varphi _4 (\mu _1, \mu _2, \mu _3, \mu _4)$ is an symmetric function of its arguments and is given by
\begin{equation}
\label{phi-4}
\begin{split}
\varphi _4 (\mu _1, \mu _2, \mu _3, \mu _4) &= f(\mu _1) f(\mu _2) f(\mu _3) f(\mu _4) \Omega _+ 
+ c_4^{(1)}( \mu _4 ; \mu _1 , \mu _2, \mu _3) f(\mu _1) f(\mu _2) f(\mu _3) \Omega _+  \\
&+ c_4^{(1)}(\mu _3 ; \mu _1, \mu _2, \mu _4) f(\mu _1) f(\mu _2) f(\mu _4) \Omega _+
+ c_4^{(1)}(\mu _2 ; \mu _1, \mu _3, \mu _4, ) f(\mu _1) f(\mu _3) f(\mu _4) \Omega _+ \\
&+ c_4^{(1)}( \mu _1 ; \mu _2 , \mu _3, \mu _4) f(\mu _2) f(\mu _3) f(\mu _4) \Omega _+
+ c_4^{(2)}(\mu _3 , \mu _4 ; \mu _1, \mu _2) f(\mu _1) f(\mu _2) \Omega _+ \\
&+ c_4^{(2)}(\mu _2 , \mu _4 ; \mu _1, \mu _3) f(\mu _1) f(\mu _3) \Omega _+
+ c_4^{(2)}(\mu _2 , \mu _3 ; \mu _1, \mu _4) f(\mu _1) f(\mu _4) \Omega _+ \\
&+ c_4^{(2)}(\mu _1 , \mu _4 ; \mu _2, \mu _3) f(\mu _2) f(\mu _3) \Omega _+
+ c_4^{(2)}(\mu _1 , \mu _3 ; \mu _2, \mu _4) f(\mu _2) f(\mu _4) \Omega _+ \\
&+ c_4^{(2)}(\mu _1 , \mu _2 ; \mu _3, \mu _4) f(\mu _3) f(\mu _4) \Omega _+
+ c_4^{(3)}(\mu _2 , \mu _3 , \mu _4 ; \mu _1) f(\mu _1) \Omega _+ \\
&+ c_4^{(3)}(\mu _1 , \mu _2 , \mu _4 ; \mu _2) f(\mu _2) \Omega _+ 
+ c_4^{(3)}(\mu _1 , \mu _2 , \mu _4 ; \mu _3) f(\mu _3) \Omega _+ \\
&+ c_4^{(3)}(\mu _1 , \mu _2 , \mu _3 ; \mu _4) f(\mu _4) \Omega _+
+ c_4^{(4)}(\mu _1, \mu _2,  \mu _3, \mu _4 ) \Omega _+,
\end{split}
\end{equation}
where the four scalar coefficients are
\begin{align}
\label{c4-1}
c_4^{(1)}( \mu _1 ; \mu _2 , \mu _3, \mu _4) &= - \frac{\psi \nu}{\xi} \left( 1 - \mu_1^2 \rho(\mu_1) + \sum _{i=2}^4 \frac{2\mu_1^2 }{\mu_1^2 - \mu_i^2} 
\right) , \\[2ex]
\label{c4-2}
c_4^{(2)}(\mu _1 , \mu _2 ; \mu _3, \mu _4)  &= - \frac{\psi ^2}{\nu ^2} \left( \frac{\xi ^2 - 3 \mu_2 ^2 \nu ^2}{\mu_1^2 - \mu_2^2} \left( \rho (\mu _1) - \sum _{i=3}^4 \frac{2}{\mu_1 ^2 - \mu _i^2} \right) 
- \frac{\xi ^2 - 3 \mu_1 ^2 \nu ^2}{\mu_1^2 - \mu_2^2} \left( \rho (\mu _2) - \sum _{j=3}^4 \frac{2}{\mu_2 ^2 - \mu _j^2} \right) \right) 	\notag \\
&- \frac{\psi ^2}{\nu ^2} (\xi ^2 - ( \mu_1 ^2 + \mu_2 ^2 ) \nu ^2) \left( \rho (\mu _1) - \sum _{i=3}^4 \frac{2}{\mu_1 ^2 - \mu _i^2} \right) \left( \rho (\mu _2) - \sum _{j=3}^4 \frac{2}{\mu_2 ^2 - \mu _j^2} \right) ,
\end{align}
\begin{align}
\label{c4-3}
&c_4^{(3)}(\mu _1 , \mu _2 , \mu _3 ; \mu _4)  = - \frac{\psi^3}{\nu ^3 \xi}	 
\left( \frac{4 \xi ^4 +  \left( \xi ^2 + \mu _1^2 \nu ^2 \right) \left( 4 \mu _1^2 - 5 ( \mu _2^2 +  \mu _3^2) \right) \nu ^2}{(\mu _1 ^2 - \mu _2 ^2)(\mu _1 ^2 - \mu _3 ^2)} \left( \rho (\mu _1) - \frac{2}{\mu_1 ^2 - \mu _4^2} \right) \right. \notag \\
&+ \frac{4 \xi ^4 + \left( \xi ^2 + \mu _2^2 \nu ^2 \right) \left( 4 \mu _2^2 - 5 ( \mu _3^2 +  \mu _1^2) \right) \nu ^2}{(\mu _2 ^2 - \mu _3 ^2)(\mu _2 ^2 - \mu _1 ^2)} \left( \rho (\mu _2) - \frac{2}{\mu_2 ^2 - \mu _4^2} \right)  \notag \\
&\left. + \frac{4 \xi ^4 + \left( \xi ^2 + \mu _3^2 \nu ^2 \right) \left( 4 \mu _3^2 - 5 ( \mu _1^2 +  \mu _2^2) \right) \nu ^2}{(\mu _3 ^2 - \mu _1 ^2)(\mu _3 ^2 - \mu _2 ^2)} \left( \rho (\mu _3) - \frac{2}{\mu_3 ^2 - \mu _4^2}\right) 
\right)	\notag 
\end{align}

\begin{align}
&- \frac{\psi^3}{\nu ^3 \xi}	\left( \frac{\xi ^2 \nu ^2 \left( \mu _1 ^4 + \mu _2 ^4 - \mu _1 ^2 \mu _2 ^2 + 2 \mu _3^2 \left( \mu _1 ^2 + \mu _2 ^2 \right) - 5 \mu _3^4 \right)
- \left( 2 \xi ^4  - \mu _1 ^2 \mu _2 ^2 \nu ^4 \right) \left( \mu _1 ^2 + \mu _2 ^2 - 2 \mu _3^2 \right)}
{(\mu _1 ^2 - \mu _3 ^2)(\mu _2 ^2 - \mu _3 ^2)} \times \right. \notag \\
&\times \left( \rho (\mu _1) - \frac{2}{\mu_1 ^2 - \mu _4^2} \right) \left( \rho (\mu _2) - \frac{2}{\mu_2 ^2 - \mu _4^2} \right)  \notag \\
&+ \frac{\xi ^2 \nu ^2 \left( \mu _2 ^4 + \mu _3 ^4 - \mu _2 ^2 \mu _3 ^2 + 2 \mu _1^2 \left( \mu _2 ^2 + \mu _3 ^2 \right) - 5 \mu _1^4 \right)
- \left( 2 \xi ^4  - \mu _2 ^2 \mu _3 ^2 \nu ^4 \right) \left( \mu _2 ^2 + \mu _3 ^2 - 2 \mu _1^2 \right)}
{(\mu _2 ^2 - \mu _1 ^2)(\mu _3 ^2 - \mu _1 ^2)} \times \notag \\
&\times \left( \rho (\mu _2) - \frac{2}{\mu_2 ^2 - \mu _4^2} \right) \left( \rho (\mu _3) - \frac{2}{\mu_3 ^2 - \mu _4^2} \right) \notag \\
&+ \frac{\xi ^2 \nu ^2 \left( \mu _3 ^4 + \mu _1 ^4 - \mu _3 ^2 \mu _1 ^2 + 2 \mu _2^2 \left( \mu _3 ^2 + \mu _1 ^2 \right) - 5 \mu _2^4 \right)
- \left( 2 \xi ^4  - \mu _3 ^2 \mu _1 ^2 \nu ^4 \right) \left( \mu _3 ^2 + \mu _1 ^2 - 2 \mu _2^2 \right)}
{(\mu _3 ^2 - \mu _2 ^2)(\mu _1 ^2 - \mu _2 ^2)} \times \notag \\
&\left. \times \left( \rho (\mu _3) - \frac{2}{\mu_3 ^2 - \mu _4^2} \right) \left( \rho (\mu _1) - \frac{2}{\mu_1 ^2 - \mu _4^2} \right) \right)  \notag \\
&- \frac{\psi^3}{\nu ^3} \xi \left(2 \xi ^2 - \left( \mu _1 ^2 + \mu _2 ^2 +  \mu _3 ^2 \right) \nu ^2 \right)
\left( \rho (\mu _1) - \frac{2}{\mu_1 ^2 - \mu _4^2} \right) \left( \rho (\mu _2) - \frac{2}{\mu_2 ^2 - \mu _4^2} \right) \left( \rho (\mu _3) - \frac{2}{\mu_3 ^2 - \mu _4^2} \right) .
\end{align}
%
%
%
%
\begin{align}
\label{c4-4}
&c_4^{(4)}(\mu _1 , \mu _2 , \mu _3 , \mu _4)  = - \frac{2\psi ^4}{\nu^4} \left( \frac{ 9 \xi ^4 + \xi ^2 \nu ^2 \left( 27 \mu _1^2  - 7 (  \mu _2^2 + \mu _3^2 + \mu _4^2 ) \right) + 3 \mu _1^2 \nu ^4
\left( 8 \mu _1^2 - 7 (\mu _2^2 + \mu _3^2 + \mu _4^2) \right)} {(\mu _1^2 - \mu _2^2)(\mu _1^2 - \mu _3^2)(\mu _1^2 - \mu _4^2)} \rho (\mu _1)  \right. \notag \\
&+ \frac{ 9 \xi ^4 + \xi ^2 \nu ^2 \left( 27 \mu _2^2  - 7 (  \mu _1^2 + \mu _3^2 + \mu _4^2 ) \right) 
+ 3 \mu _2^2 \nu ^4 \left( 8 \mu _2^2 - 7 (\mu _2^2 + \mu _3^2 + \mu _4^2) \right)} {(\mu _2^2 - \mu _1^2)(\mu _2^2 - \mu _3^2)(\mu _2^2 - \mu _4^2)} \rho (\mu _2)  \notag \\
&+ \frac{ 9 \xi ^4 + \xi ^2 \nu ^2 \left( 27 \mu _3^2  - 7 (  \mu _1^2 + \mu _2^2 + \mu _4^2 ) \right) 
+ 3 \mu _3^2 \nu ^4 \left( 8 \mu _3^2 - 7 (\mu _1^2 + \mu _2^2 + \mu _4^2) \right)} {(\mu _3^2 - \mu _1^2)(\mu _3^2 - \mu _2^2)(\mu _3^2 - \mu _4^2)} \rho (\mu _3)
\notag \\
&\left. + \frac{ 9 \xi ^4 + \xi ^2 \nu ^2 \left( 27 \mu _4^2  - 7 (  \mu _1^2 + \mu _2^2 + \mu _3^2 ) \right) 
+ 3 \mu _4^2 \nu ^4 \left( 8 \mu _4^2 - 7 (\mu _1^2 + \mu _2^2 + \mu _3^2) \right)} {(\mu _4^2 - \mu _1^2)(\mu _4^2 - \mu _2^2)(\mu _4^2 - \mu _3^2)} \rho (\mu _4)  \right) \notag \\[1.5ex]
&- \frac{\psi ^4}{\nu^4}    \left( 3 \xi ^4 \left( 2 ( \mu _1^4 +  \mu _1^2 \mu _2^2 + \mu _2^4) - 3 ( \mu _1^2
\mu _3^2 + \mu _1^2 \mu _4^2 + \mu _2^2 \mu _3^2 + \mu _2^2 \mu _4^2 - 2 \mu _3^2 \mu _4^2 )
\right) -18 \xi ^2 \nu ^2  \mu _1^2 \mu _2^2 (\mu _3^2 + \mu _4^2) \right. \notag \\
&+\xi ^2 \nu ^2 \left( -2 (\mu_1^6 - 6 \mu_1^4 \mu_2^2 -6 \mu_1^2 \mu_2^4 + \mu _2^6) - 4 (\mu _1^2 + \mu _2^2)^2 (\mu _3^2 + \mu _4^2) + 7 (\mu _1^2 + \mu _2^2) (\mu _3^2 + \mu _4^2)^2 + 2 \mu_3^2 \mu_4^2 \times \right. \notag \\
&\left. \times (5 (\mu _1^2 + \mu _2^2) - 7 (\mu _3^2 + \mu _4^2) ) \right) + \nu ^4 \left( 14 \mu_1^2 \mu_2^2 (\mu_3^4 + \mu_4^4) - (4 \mu_1^2 \mu_2^2 + 7 \mu_3^2 \mu_4^2) (\mu_1^2 + \mu_2^2) (\mu_3^2 + \mu_4^2) \right) + 2\nu ^4 \times \notag \\
&\left. \times \left(  4 \mu_3^2 \mu_4^2  (\mu_1^2 + \mu_2^2) ^2 - 3 \mu_1^2 \mu_2^2  (\mu_1^2 - \mu_2^2) ^2 - \mu_1^2 \mu_2^2 ( 3 \mu_1^2 \mu_2^2 + 5 \mu_3^2 \mu_4^2) \right) \right) 
\frac{\rho (\mu _1) \rho (\mu _2)}{(\mu _1^2 - \mu _3^2)(\mu _1^2 - \mu _4^2)(\mu _2^2 - \mu _3^2)(\mu _2^2 - \mu _4^2)}  \notag 
\end{align}
\begin{align}
&- \frac{\psi ^4}{\nu^4}    \left( 3 \xi ^4 \left( 2 ( \mu _1^4 +  \mu _1^2 \mu _3^2 + \mu _3^4) - 3 ( \mu _1^2
\mu _2^2 + \mu _1^2 \mu _4^2 + \mu _2^2 \mu _3^2 + \mu _3^2 \mu _4^2 - 2 \mu _2^2 \mu _4^2 )
\right) -18 \xi ^2 \nu ^2  \mu _1^2 \mu _3^2 (\mu _2^2 + \mu _4^2) \right. \notag \\
&+ \xi ^2 \nu ^2 \left( -2 (\mu_1^6 - 6 \mu_1^4 \mu_3^2 -6 \mu_1^2 \mu_3^4 + \mu _3^6) - 4 (\mu _1^2 + \mu _3^2)^2 (\mu _2^2 + \mu _4^2) + 7 (\mu _1^2 + \mu _3^2) (\mu _2^2 + \mu _4^2)^2 + 2 \mu_2^2 \mu_4^2 \times \right. \notag \\
&\left. \times (5 (\mu _1^2 + \mu _3^2) - 7 (\mu _2^2 + \mu _4^2) ) \right) + \nu ^4 \left( 14 \mu_1^2 \mu_3^2 (\mu_2^4 + \mu_4^4) - (4 \mu_1^2 \mu_3^2 + 7 \mu_2^2 \mu_4^2) (\mu_1^2 + \mu_3^2) (\mu_2^2 + \mu_4^2) \right) + 2\nu ^4 \times \notag \\
&\left. \times \left( 4 \mu_2^2 \mu_4^2  (\mu_1^2 + \mu_3^2) ^2 - 3 \mu_1^2 \mu_3^2  (\mu_1^2 - \mu_3^2) ^2 - \mu_1^2 \mu_3^2 ( 3 \mu_1^2 \mu_3^2 + 5 \mu_2^2 \mu_4^2) \right) \right) 
\frac{\rho (\mu _1) \rho (\mu _3)}{(\mu _1^2 - \mu _2^2)(\mu _1^2 - \mu _4^2)(\mu _3^2 - \mu _2^2)(\mu _3^2 - \mu _4^2)}  \notag \\[1.5ex]
&- \frac{\psi ^4}{\nu^4}    \left( 3 \xi ^4 \left( 2 ( \mu _1^4 +  \mu _1^2 \mu _4^2 + \mu _4^4) - 3 ( \mu _1^2
\mu _2^2 + \mu _1^2 \mu _3^2 + \mu _2^2 \mu _4^2 + \mu _3^2 \mu _4^2 - 2 \mu _2^2 \mu _3^2 )
\right) -18 \xi ^2 \nu ^2  \mu _1^2 \mu _4^2 (\mu _2^2 + \mu _3^2) \right. \notag \\
&+ \xi ^2 \nu ^2 \left( -2 (\mu_1^6 - 6 \mu_1^4 \mu_4^2 -6 \mu_1^2 \mu_4^4 + \mu _4^6) - 4 (\mu _1^2 + \mu _4^2)^2 (\mu _2^2 + \mu _3^2) + 7 (\mu _1^2 + \mu _4^2) (\mu _2^2 + \mu _3^2)^2 + 2 \mu_2^2 \mu_3^2 \times \right. \notag \\
&\left. \times (5 (\mu _1^2 + \mu _4^2) - 7 (\mu _2^2 + \mu _3^2) ) \right) + \nu ^4 \left( 14 \mu_1^2 \mu_4^2 (\mu_2^4 + \mu_3^4) - (4 \mu_1^2 \mu_4^2 + 7 \mu_2^2 \mu_3^2) (\mu_1^2 + \mu_4^2) (\mu_2^2 + \mu_3^2) \right) + 2 \nu ^4 \times \notag \\
&\left. \times \left( 4 \mu_2^2 \mu_3^2  (\mu_1^2 + \mu_4^2) ^2 - 3 \mu_1^2 \mu_4^2  (\mu_1^2 - \mu_4^2) ^2 - \mu_1^2 \mu_4^2 ( 3 \mu_1^2 \mu_4^2 + 5 \mu_2^2 \mu_3^2) \right) \right) 
\frac{\rho (\mu _1) \rho (\mu _4)}{(\mu _1^2 - \mu _2^2)(\mu _1^2 - \mu _3^2)(\mu _4^2 - \mu _2^2)(\mu _4^2 - \mu _3^2)}  \notag \\[1.5ex]
&- \frac{\psi ^4}{\nu^4}  \left( 3 \xi ^4 \left( 2 ( \mu _2^4 +  \mu _2^2 \mu _3^2 + \mu _3^4) - 3 ( \mu _1^2
\mu _2^2 + \mu _2^2 \mu _4^2 + \mu _1^2 \mu _3^2 + \mu _3^2 \mu _4^2 - 2 \mu _1^2 \mu _4^2 )
\right) -18 \xi ^2 \nu ^2  \mu _2^2 \mu _3^2 (\mu _1^2 + \mu _4^2) \right. \notag \\
&+ \xi ^2 \nu ^2 \left( -2 (\mu_2^6 - 6 \mu_2^4 \mu_3^2 -6 \mu_2^2 \mu_3^4 + \mu _3^6) - 4 (\mu _2^2 + \mu _3^2)^2 (\mu _1^2 + \mu _4^2) + 7 (\mu _2^2 + \mu _3^2) (\mu _1^2 + \mu _4^2)^2 + 2 \mu_1^2 \mu_4^2 \times \right. \notag \\
&\left. \times (5 (\mu _2^2 + \mu _3^2) - 7 (\mu _1^2 + \mu _4^2) ) \right) + \nu ^4 \left( 14 \mu_2^2 \mu_3^2 (\mu_1^4 + \mu_4^4) - (4 \mu_2^2 \mu_3^2 + 7 \mu_1^2 \mu_4^2) (\mu_2^2 + \mu_3^2) (\mu_1^2 + \mu_4^2) \right) + 2 \nu ^4 \times \notag \\
&\left. \times \left( 4 \mu_1^2 \mu_4^2  (\mu_2^2 + \mu_3^2) ^2 - 3 \mu_2^2 \mu_3^2  (\mu_2^2 - \mu_3^2) ^2 - \mu_2^2 \mu_3^2 ( 3 \mu_2^2 \mu_3^2 + 5 \mu_1^2 \mu_4^2) \right) \right) 
\frac{\rho (\mu _2) \rho (\mu _3)}{(\mu _2^2 - \mu _1^2)(\mu _2^2 - \mu _4^2)(\mu _3^2 - \mu _1^2)(\mu _3^2 - \mu _4^2)}  \notag 
\end{align}
\begin{align}
&- \frac{\psi ^4}{\nu^4} \left( 3 \xi ^4 \left( 2 ( \mu _2^4 +  \mu _2^2 \mu _4^2 + \mu _4^4) - 3 ( \mu _1^2
\mu _2^2 + \mu _2^2 \mu _3^2 + \mu _1^2 \mu _4^2 + \mu _3^2 \mu _4^2 - 2 \mu _1^2 \mu _3^2 )
\right) -18 \xi ^2 \nu ^2  \mu _2^2 \mu _4^2 (\mu _1^2 + \mu _3^2) \right. \notag \\
&+ \xi ^2 \nu ^2 \left( -2 (\mu_2^6 - 6 \mu_2^4 \mu_4^2 -6 \mu_2^2 \mu_4^4 + \mu _4^6) - 4 (\mu _2^2 + \mu _4^2)^2 (\mu _1^2 + \mu _3^2) + 7 (\mu _2^2 + \mu _4^2) (\mu _1^2 + \mu _3^2)^2 + 2 \mu_1^2 \mu_3^2 \times \right. \notag \\
&\left. \times (5 (\mu _2^2 + \mu _4^2) - 7 (\mu _1^2 + \mu _3^2) ) \right) + \nu ^4 \left( 14 \mu_2^2 \mu_4^2 (\mu_1^4 + \mu_3^4) - (4 \mu_2^2 \mu_4^2 + 7 \mu_1^2 \mu_3^2) (\mu_2^2 + \mu_4^2) (\mu_1^2 + \mu_3^2) \right) + 2 \nu ^4 \times \notag \\
&\left. \times \left( 4 \mu_1^2 \mu_3^2  (\mu_2^2 + \mu_4^2) ^2 - 3 \mu_2^2 \mu_4^2  (\mu_2^2 - \mu_4^2) ^2 - \mu_2^2 \mu_4^2 ( 3 \mu_2^2 \mu_4^2 + 5 \mu_1^2 \mu_3^2) \right) \right) 
\frac{\rho (\mu _2) \rho (\mu _4)}{(\mu _2^2 - \mu _1^2)(\mu _2^2 - \mu _3^2)(\mu _4^2 - \mu _1^2)(\mu _4^2 - \mu _3^2)}  \notag \\[1.5ex]
&- \frac{\psi ^4}{\nu^4} \left( 3 \xi ^4 \left( 2 ( \mu _3^4 +  \mu _3^2 \mu _4^2 + \mu _4^4) - 3 ( \mu _2^2
\mu _3^2 + \mu _1^2 \mu _3^2 + \mu _2^2 \mu _4^2 + \mu _1^2 \mu _4^2 - 2 \mu _1^2 \mu _2^2 )
\right) -18 \xi ^2 \nu ^2  \mu _3^2 \mu _4^2 (\mu _1^2 + \mu _2^2) \right. \notag \\
&+ \xi ^2 \nu ^2 \left( -2 (\mu_3^6 - 6 \mu_3^4 \mu_4^2 -6 \mu_3^2 \mu_4^4 + \mu _4^6) - 4 (\mu _3^2 + \mu _4^2)^2 (\mu _1^2 + \mu _2^2) + 7 (\mu _3^2 + \mu _4^2) (\mu _1^2 + \mu _2^2)^2 + 2 \mu_1^2 \mu_2^2 \times \right. \notag \\
&\left. \times (5 (\mu _3^2 + \mu _4^2) - 7 (\mu _1^2 + \mu _2^2) ) \right) + \nu ^4 \left( 14 \mu_3^2 \mu_4^2 (\mu_1^4 + \mu_2^4) - (4 \mu_3^2 \mu_4^2 + 7 \mu_1^2 \mu_2^2) (\mu_3^2 + \mu_4^2) (\mu_1^2 + \mu_2^2) \right) + 2 \nu ^4 \times \notag \\
&\left. \times \left( 4 \mu_1^2 \mu_2^2  (\mu_3^2 + \mu_4^2) ^2 - 3 \mu_3^2 \mu_4^2  (\mu_3^2 - \mu_4^2) ^2 - \mu_3^2 \mu_4^2 ( 3 \mu_3^2 \mu_4^2 + 5 \mu_1^2 \mu_2^2) \right) \right) 
\frac{\rho (\mu _3) \rho (\mu _4)}{(\mu _3^2 - \mu _1^2)(\mu _3^2 - \mu _2^2)(\mu _4^2 - \mu _1^2)(\mu _4^2 - \mu _2^2)}  \notag 
\end{align}
\begin{align}
&- \frac{\psi ^4}{\nu^4}   \left( 3 \xi ^4 \left( 3 \mu _4 ^4 - 2 \mu _4^2 (\mu _1^2 + \mu _2^2 + \mu _3^2) + \mu _1^2 \mu _2^2 + \mu _1^2 \mu _3^2 + \mu _2^2 \mu _3^2 \right) - \xi ^2 \nu ^2 \left( 7 \mu_4 ^6 - 4  \mu_4 ^4 (\mu _1^2 + \mu _2^2 + \mu _3^2) + \mu_4 ^2  \times
\right. \right. \notag \\
&\left. \times (3 (\mu _1^2 \mu _2^2 + \mu _1^2 \mu _3^2 + \mu _2^2 \mu _3^2 ) - 2(\mu _1^4 + \mu _2^4 + \mu _3^4) ) 
+ \mu _1^4 \mu _2^2 + \mu _1^4 \mu _3^2 + \mu _1^2 \mu _2^4 + \mu _1^2 \mu _3^4 + \mu _2^4 \mu _3^2 + \mu _2^2 \mu _3^4 - 4 \mu _1^2 \mu _2^2 \mu _3^2 \right)
\notag \\
&- \nu ^4 \left( 2 \mu _4 ^4 (\mu_1^2 \mu_2^2 + \mu_1^2 \mu_3^2 + \mu_2^2 \mu_3^2) 
- \mu_4 ^2 ( \mu_1^4 \mu_2^2 +  \mu_1^2 \mu_2^4 + \mu_1^4 \mu_3^2 + \mu_1^2 \mu_3^4 + \mu_2^4 \mu_3^2 + \mu_2^2 \mu_3^4 + 6 \mu_1^2 \mu_2^2 \mu_3^2) \right)
\notag \\
&\left. - \nu ^4 \left( 2  \mu _1^2 \mu _2^2 \mu _3^2 (\mu _1^2 + \mu _2^2 + \mu _3^2) \right) \right) 
\frac{\rho (\mu _1) \rho (\mu _2) \rho (\mu _3)}{(\mu _1^2 - \mu _4^2)(\mu _2^2 - \mu _4^2)(\mu _3^2 - \mu _4^2)}  \notag \\[1.5ex]
&- \frac{\psi ^4}{\nu^4}   \left( 3 \xi ^4 \left( 3 \mu _3^4 - 2 \mu _3^2 (\mu _1^2 + \mu _2^2 + \mu _4^2) + \mu _1^2 \mu _2^2 + \mu _1^2 \mu _4^2 + \mu _2^2 \mu _4^2 \right) - \xi ^2 \nu ^2 \left( 7 \mu_3 ^6 - 4  \mu_3 ^4 (\mu _1^2 + \mu _2^2 + \mu _4^2) + \mu_3 ^2  \times
\right. \right. \notag \\
&\left. \times (3 (\mu _1^2 \mu _2^2 + \mu _1^2 \mu _4^2 + \mu _2^2 \mu _4^2 ) - 2(\mu _1^4 + \mu _2^4 + \mu _4^4) ) + \mu _1^4 \mu _2^2 + \mu _1^4 \mu _4^2 + \mu _1^2 \mu _2^4 + \mu _1^2 \mu _4^4 + \mu _2^4 \mu _4^2 + \mu _2^2 \mu _4^4 - 4 \mu _1^2 \mu _2^2 \mu _4^2 \right)
\notag \\
&- \nu ^4 \left( 2 \mu _3 ^4 (\mu_1^2 \mu_2^2 + \mu_1^2 \mu_4^2 + \mu_2^2 \mu_4^2) 
- \mu_3 ^2 ( \mu_1^4 \mu_2^2 +  \mu_1^2 \mu_2^4 + \mu_1^4 \mu_4^2 + \mu_1^2 \mu_4^4 + \mu_2^4 \mu_4^2 + \mu_2^2 \mu_4^4 + 6 \mu_1^2 \mu_2^2 \mu_4^2) \right)
\notag \\
&\left. - \nu ^4 \left( 2  \mu _1^2 \mu _2^2 \mu _4^2 (\mu _1^2 + \mu _2^2 + \mu _4^2) \right) \right) 
\frac{\rho (\mu _1) \rho (\mu _2) \rho (\mu _4)}{(\mu _1^2 - \mu _3^2)(\mu _2^2 - \mu _3^2)(\mu _4^2 - \mu _3^2)}  \notag \\[1.5ex]
&- \frac{\psi ^4}{\nu^4}   \left( 3 \xi ^4 \left( 3 \mu _2^4 - 2 \mu _2^2 (\mu _1^2 + \mu _3^2 + \mu _4^2) + \mu _1^2 \mu _3^2 + \mu _1^3 \mu _4^2 + \mu _3^2 \mu _4^2 \right) - \xi ^2 \nu ^2 \left( 7 \mu_2 ^6 - 4  \mu_2 ^4 (\mu _1^2 + \mu _3^2 + \mu _4^2) + \mu_2 ^2  \times
\right. \right. \notag \\
&\left. \times (3 (\mu _1^2 \mu _3^2 + \mu _1^2 \mu _4^2 + \mu _3^2 \mu _4^2 ) - 2(\mu _1^4 + \mu _3^4 + \mu _4^4) ) + \mu _1^4 \mu _3^2 + \mu _1^4 \mu _4^2 + \mu _1^2 \mu _3^4 + \mu _1^2 \mu _4^4 + \mu _3^4 \mu _4^2 + \mu _3^2 \mu _4^4 - 4 \mu _1^2 \mu _3^2 \mu _4^2 \right)
\notag \\
&- \nu ^4 \left( 2 \mu _2 ^4 (\mu_1^2 \mu_3^2 + \mu_1^2 \mu_4^2 + \mu_3^2 \mu_4^2) 
- \mu_2 ^2 ( \mu_1^4 \mu_3^2 +  \mu_1^2 \mu_3^4 + \mu_1^4 \mu_4^2 + \mu_1^2 \mu_4^4 + \mu_3^4 \mu_4^2 + \mu_3^2 \mu_4^4 + 6 \mu_1^2 \mu_3^2 \mu_4^2) \right)
\notag \\
&\left. - \nu ^4 \left( 2  \mu _1^2 \mu _3^2 \mu _4^2 (\mu _1^2 + \mu _3^2 + \mu _4^2) \right) \right) 
\frac{\rho (\mu _1) \rho (\mu _3) \rho (\mu _4)}{(\mu _1^2 - \mu _2^2)(\mu _3^2 - \mu _2^2)(\mu _4^2 - \mu _2^2)}  \notag 
\end{align}
\begin{align}
&- \frac{\psi ^4}{\nu^4}   \left( 3 \xi ^4 \left( 3 \mu _1^4 - 2 \mu _1^2 (\mu _2^2 + \mu _3^2 + \mu _4^2) + \mu _2^2 \mu _3^2 + \mu _2^3 \mu _4^2 + \mu _3^2 \mu _4^2 \right) - \xi ^2 \nu ^2 \left( 7 \mu_1 ^6 - 4  \mu_1 ^4 (\mu _2^2 + \mu _3^2 + \mu _4^2) + \mu_1 ^2  \times
\right. \right. \notag \\
&\left. \times (3 (\mu _2^2 \mu _3^2 + \mu _2^2 \mu _4^2 + \mu _3^2 \mu _4^2 ) - 2(\mu _2^4 + \mu _3^4 + \mu _4^4) ) + \mu _2^4 \mu _3^2 + \mu _2^4 \mu _4^2 + \mu _2^2 \mu _3^4 + \mu _2^2 \mu _4^4 + \mu _3^4 \mu _4^2 + \mu _3^2 \mu _4^4 - 4 \mu _2^2 \mu _3^2 \mu _4^2 \right)
\notag \\
&- \nu ^4 \left( 2 \mu _1 ^4 (\mu_2^2 \mu_3^2 + \mu_2^2 \mu_4^2 + \mu_3^2 \mu_4^2) 
- \mu_1 ^2 ( \mu_2^4 \mu_3^2 +  \mu_2^2 \mu_3^4 + \mu_2^4 \mu_4^2 + \mu_2^2 \mu_4^4 + \mu_3^4 \mu_4^2 + \mu_3^2 \mu_4^4 + 6 \mu_2^2 \mu_3^2 \mu_4^2) \right)
\notag \\
&\left. - \nu ^4 \left( 2  \mu _2^2 \mu _3^2 \mu _4^2 (\mu _2^2 + \mu _3^2 + \mu _4^2) \right) \right) 
\frac{\rho (\mu _2) \rho (\mu _3) \rho (\mu _4)}{(\mu _2^2 - \mu _1^2)(\mu _3^2 - \mu _1^2)(\mu _4^2 - \mu _1^2)}  \notag \\[1.5ex]
&- \frac{\psi^4}{\nu ^4} \xi ^2 \left( 3 \xi ^2 - \left( \mu _1 ^2 + \mu _2 ^2 + \mu _3 ^2 + \mu _4 ^2 \right) \nu ^2 \right) \rho (\mu _1) \rho (\mu _2) \rho (\mu _3) \rho (\mu _4)
\end{align}

\end{document}